\documentclass[a4paper,11pt]{article}

\usepackage{jcappub} 
\usepackage[T1]{fontenc} 
\usepackage{natbib}
\usepackage{amssymb}
\usepackage{amsmath}
\usepackage{amsfonts}
\usepackage{hyperref}
\usepackage{graphicx}
\usepackage[hang,tight,raggedright]{subfigure}
\usepackage{multirow}
\usepackage{subfigure}
\usepackage{float}
\usepackage{graphics}
\usepackage{slashed}
\usepackage{color}
\usepackage{bm}
\usepackage{braket}
\usepackage{psfrag}
\usepackage{booktabs}
\usepackage{latexsym}
\usepackage{pythonhighlight}
\usepackage{mathrsfs}

\newcommand{\be}{\begin{equation}}
\newcommand{\ee}{\end{equation}}
\newcommand{\besp}{\begin{equation}\begin{split}}
\newcommand{\eesp}{\end{split}\end{equation}}

\title{\boldmath \texttt{DeepSSM}: an emulator of gravitational wave spectra from sound waves during cosmological first-order phase transitions}

\author[a]{Chi Tian,}
\author[b,1]{Xiao Wang\note{Corresponding author.},}
\author[b]{and Csaba Bal\'azs}

\affiliation[a]{School of Physics and Optoelectronics Engineering, Anhui University, 111 Jiulong Road, Hefei, Anhui, China 230601}
\affiliation[b]{School of Physics and Astronomy, Monash University, Melbourne 3800 Victoria, Australia}

\emailAdd{ctian@ahu.edu.cn}
\emailAdd{xiao.wang1@monash.edu}
\emailAdd{csaba.balazs@monash.edu}

\abstract{We present \texttt{DeepSSM}, an open-source code powered by neural networks (NNs) to emulate gravitational wave (GW) spectra produced by sound waves during cosmological first-order phase transitions in the radiation-dominated era. 
 The training data is obtained from an enhanced version of the Sound Shell Model (SSM), which accounts for the effects of cosmic expansion and yields more accurate spectra in the infrared regime.  The emulator enables instantaneous predictions of GW spectra given the phase transition parameters, while achieving agreement with the enhanced SSM model within 
10\% accuracy in the worst-case scenarios. The emulator is highly computationally efficient and fully differentiable, making it particularly suitable for direct Bayesian inference on phase transition parameters without relying on empirical templates, such as broken power-law models. We demonstrate this capability by successfully reconstructing phase transition parameters and their degeneracies  from mock LISA observations using a Hamiltonian Monte Carlo sampler. The code is available at: \url{https://github.com/ctian282/DeepSSM}.}

\begin{document}
\maketitle
\flushbottom

\section{Introduction}
\label{sec:intro}

Recent observations from pulsar timing arrays in the nano-hertz range \cite{NANOGrav:2023gor, Xu:2023wog, EPTA:2023fyk, Reardon:2023gzh, Miles:2024seg} have provided preliminary evidence for the existence of a stochastic gravitational wave background (SGWB) and have offered initial insights into the shape of its power spectrum.
The next generation of space- and ground-based gravitational wave observatories \cite{LIGOScientific:2016jlg, TianQin:2015yph,Hu:2017mde, Ruan:2018tsw, Kawamura:2011zz, Maggiore:2019uih} are expected to thoroughly characterise this stochastic signal across different frequency bands.  The significant interest in this new observable arises from its connection not only to astrophysical phenomena but also to various complex physical processes in the early Universe.  Many of these processes are closely related to physics beyond the Standard Model (BSM) of particle physics \cite{Boyanovsky:2006bf, Mazumdar:2018dfl, Athron:2023xlk}, highlighting the critical importance of precise SGWB measurements for exploring potential new physics.

Cosmological first-order phase transitions (FOPTs) are some of the most important physical processes that occur during the radiation-dominated era.  During these transitions, GWs can be generated through three primary mechanisms: bubble collisions, sound waves, and turbulence.  These mechanisms collectively produce stochastic GWs, which constitute a significant component of the SGWB.  Among these production channels, recent studies~\cite{Caprini:2015zlo,Caprini:2019egz,Athron:2023xlk} suggest that sound waves generally serve as the dominant source of GWs in the case of thermal FOPTs.  Models quantifying the resulting GW spectra from the FOPTs are typically based on some phase transition parameters, such as the strength, the duration, the characteristic temperature of the transition, and the bubble wall velocity.  Notably, these parameters also depend on the particular extension of the Standard Model, as shown by \cite{Friedrich:2022cak, Freitas:2021yng, Cline:2021iff, Paul:2020wbz, Niemi:2020hto, Wang:2019pet, Hashino:2018wee,Tian:2024ysd,Wang:2024slx}. Therefore, phase transition parameters serve as crucial links between new physics and the SGWB, which is a physical observable.

To reveal how the resulting GW spectra depend on these phase transition parameters, sophisticated scalar+fluid lattice simulations~\cite{Hindmarsh:2013xza,Hindmarsh:2015qta,Hindmarsh:2017gnf} have been developed.  These simulations require a large volume to accommodate hundreds of bubbles and fine grid spacing to resolve the bubble wall thickness, making them numerically expensive and time-consuming.  To address these challenges, a simplified semi-analytical framework, the Sound Shell Model (SSM)~\cite{Hindmarsh:2016lnk, Hindmarsh:2019phv, Guo:2020grp, Wang:2021dwl, Cai:2023guc, RoperPol:2023dzg, Giombi:2024kju}, was originally proposed in~\cite{Hindmarsh:2016lnk,Hindmarsh:2019phv}.  This model enables the prediction of the SGWB spectra on the basis of a given set of phenomenological parameters in a more convenient manner.  Although the SSM typically yields satisfactory predictions at higher frequencies, it often produces a steep rise in the spectrum at lower frequencies, scaling as $\Omega_{\mathrm{GW}} \sim k^9$, in contrast to many numerical studies~\cite{Hindmarsh:2013xza,Hindmarsh:2015qta,Jinno:2022mie}, which suggest a shallower scaling of $\sim k^3$ in the low-frequency regime.  To resolve this discrepancy, the authors of ref. \cite{RoperPol:2023dzg} revised the traditional SSM by examining the validity of various approximations, resulting in an SGWB spectrum with more complicated and accurate low-frequency behaviour.  This modification provides a more robust and reliable semi-analytical template for studying the GW spectra.

Although these semi-analytical models offer a convenient framework for predicting SGWB spectra, they typically require high-precision, multi-dimensional numerical integrations.  This results in significant computational overhead, making them impractical to be applied to some analysis pipelines of the SGWB data.
In practice, studies such as \cite{Caprini:2019pxz, Flauger:2020qyi, Giese:2021dnw, Gowling:2022pzb} often resort to empirical templates, such as broken or double-broken power-law models, to perform Bayesian inference, even though there are discrepancies between the results of the semi-analytical model and these power-law models \cite{Guo:2024gmu}.  The situation becomes even more challenging when considering the enhanced SSM, as it introduces additional phase transition parameters and higher-dimensional integrations, further adding computational complexity while producing spectra deviating more significantly from a power-law shape.

Given the necessity of a highly efficient model for computing the GW spectra, developing an emulator for gravitational waves sourced by phase transitions becomes crucial.
The goal of this emulator is to efficiently map a set of phase transition parameters to the corresponding GW spectrum.  Since this is a highly non-linear projection, neural networks (NNs) present a particularly promising solution.  NNs are well-established for their ability to approximate any smooth function \cite{cybenko1989approximation, HORNIK1989359, HORNIK1991251, LESHNO1993861}, making them particularly well-suited to build such emulators.  In recent years, NN-based emulators have been developed for computing various cosmological signals, including those from the cosmic microwave background \cite{Auld:2006pm, Auld:2007qz, SpurioMancini:2021ppk, Pal:2022hpi, Nygaard:2022wri, Gunther:2022pto, Bonici:2023xjk} (including temperature, polarisation, and lensing), large-scale structures \cite{Agarwal:2012ew, Agarwal:2013aea, Manrique-Yus:2019hqc, Angulo:2020vky, Arico:2021izc, DeRose:2021pqx, Bose:2022vwi, Bonici:2025ltp}, and 21-cm signals \cite{Kern:2017ccn, Schmit:2017pho, Bevins:2021eah}.  A similar framework could thus be adopted to predict the SGWB spectra arising from cosmological first-order phase transitions.

In this paper, we present an emulator \texttt{DeepSSM} for GW spectra from cosmological phase transitions, which is based on a deep-neuron network trained on data generated by the enhanced SSM.  This emulator produces very accurate estimations while achieving several orders of magnitude improvement in computational speed compared to traditional semi-analytical approaches, even without the help of graphics processing units (GPUs). This makes it particularly well-suited for Bayesian estimation of phase transition parameters.  To demonstrate the applications of our model for parameter estimation, we inject mock GW spectra with the realistic LISA noise and successfully employ the Hamiltonian Monte Carlo (HMC) method to reconstruct the phase transition parameters from the simulated data.  Our results highlight the potential utility of this emulator for data analysis in both ongoing and forthcoming experimental efforts.

The structure of this paper is as follows.  In Section~\ref{sec:SSM}, we present a brief review of the SSM and its modifications that give a more reasonable low-frequency spectrum.  In Section~\ref{sec:NN}, the architecture of the neural network used in our emulator is described in detail, along with the training and validation strategies employed. Bayesian inference as a demonstration for possible applications of our model is presented in Section~\ref{sec:LISA}.  We discuss future improvements on \texttt{DeepSSM} and conclude in Section~\ref{sec:conc}.

\section{GW production from sound waves}
\label{sec:SSM}

GWs can be characterised by tensor-mode perturbations $h_{ij}$ of the metric.  
The metric of the Friedmann-Lema\^itre-Robertson-Walker Universe in conformal coordinates is given by
\begin{equation}
    ds^2 = a^2(\tau)[-d\tau^2 + (\delta_{ij} + h_{ij} )dx^idx^j], 
\end{equation}
where $a(\tau)$ is the scale factor and $\tau$ is the conformal time.
The time evolution of GWs can be described by the linearised Einstein equations.  During a radiation domination era these equations, in momentum space, can be expressed by
\begin{equation}
    (\partial_\tau^2 + k^2) \tilde{\ell}_{ij}(\tau, \mathbf{k}) = 16\pi G a^3 \bar{\rho}\tilde{\Pi}_{ij}(\tau, \mathbf{k}), 
    \label{eq:EE}
\end{equation}
where we have defined $\tilde{\ell}_{ij}(\tau, \mathbf{k}) \equiv h_{ij}(\tau, \mathbf{k})/a(\tau)$ and $\tilde{\Pi}_{ij}(\tau, \mathbf{k})$ is the transverse-traceless (TT) part of the anisotropic stress tensor with $G$ being the gravitational constant.  By defining the TT projection operator $\Lambda_{ij,lm} \equiv P_{il}P_{jm} - \frac{1}{2}P_{ij}P_{lm}$, with $P_{ij} \equiv \delta_{ij} -\hat{k}_i\hat{k}_j$, the term $\bar{\rho} \Pi_{ij}(\tau, \mathbf{k})$ can be expressed as
\begin{align}
    \bar{\rho} \tilde{\Pi}_{ij}(\tau, \mathbf{k}) \equiv  \Lambda_{ij,lm} T_{lm}(\tau, \mathbf{k})\,.
\end{align}
Here, $\bar{\rho} = 3\mathcal{H}^2/(8\pi G a^2)$ is the total energy density, with $\mathcal{H} \equiv a'/a$ being the conformal Hubble parameter, where $'$ denotes the derivative with respect to $\tau$.

We assume that the source of GWs is active when $\tau_* \le \tau \le \tau_{\rm fin}$, which means that gravitational waves freely propagate when $\tau > \tau_{\rm fin}$.
We further set $a(\tau_*) = 1$, so that $a(\tau) = \mathcal{H}_*\tau$, where $\mathcal{H}_* = \mathcal{H}(\tau_*)$ is the conformal Hubble parameter at $\tau_*$.
Therefore, during radiation domination, the solution of eq.~\eqref{eq:EE} can be expressed as
\begin{equation}
    \tilde{\ell}_{ij}(\tau, \mathbf{k}) = 6\mathcal{H}_*
    \begin{cases}
        {\displaystyle\int_{\tau_*}^{\tau}  \frac{d\tau_1}{\tau_1}} \tilde{\Pi}_{ij}(\tau_1, \mathbf{k}) {\displaystyle\frac{\sin k(\tau - \tau_1)}{k}}, & \tau_* <\tau <\tau_{\rm fin},\\
    &\\
    {\displaystyle\int_{\tau_*}^{\tau_{\rm fin}} \frac{d\tau_1}{\tau_1}} \tilde{\Pi}_{ij}(\tau_1, \mathbf{k}) {\displaystyle\frac{\sin k(\tau - \tau_1)}{k}}, & \tau > \tau_{\rm fin}.
    \end{cases}
\end{equation}
With these solutions, the energy density of gravitational waves is defined as
\begin{equation}
    \rho_{\rm gw} = \frac{1}{32\pi G}\langle \tilde{\ell}_{ij}^{'}\tilde{\ell}_{ij}^{*'}\rangle = \frac{1}{32\pi Ga^2}\langle h_{ij}^{'} h_{ij}^{*'}\rangle.
\end{equation}
Using the fractional gravitational wave energy density $\Omega_{\rm GW} \equiv \rho_{\rm GW}/\bar{\rho}$, 
the power spectrum of the SGWB can be written as 
\begin{equation}
    \Omega_{\rm GW} \equiv \frac{1}{\bar{\rho}}\frac{d\rho_{\rm GW}}{d\ln k}.
\end{equation}
Therefore, to derive the GW spectrum, one needs to calculate the two-point correlation function of $\tilde{\ell}_{ij}'$.
In the comoving momentum space, for $\tau_{1,2}\gg \tau_{\rm fin}$, we have 
\begin{equation}
\begin{split}
    \langle \tilde{\ell}_{ij}^{~'}(\tau_1, \mathbf{k}) \tilde{\ell}_{ij}^{*'}(\tau_2, \mathbf{k}_2) \rangle =& (6\mathcal{H}_*)^2 \int_{\tau_*}^{\tau_{\rm fin}} \frac{d\tau_1}{\tau_1} \int_{\tau_*}^{\tau_{\rm fin}} \frac{d\tau_2}{\tau_2}  \cos k(\tau - \tau_1) \cos k_2(\tau - \tau_2)\\
    &\times \Big\langle \tilde{\Pi}_{ij}(\tau_1, \mathbf{k})\tilde{\Pi}_{ij}(\tau_2, \mathbf{k}_2)\Big\rangle,
\end{split}
\end{equation}
where the unequal time correlator (UETC) of the shear stress, $U_\Pi$, is defined as
\begin{equation}
    \Big\langle \tilde{\Pi}_{ij}(\tau_1, \mathbf{k})\tilde{\Pi}_{ij}(\tau_2, \mathbf{k}_2)\Big\rangle
    \equiv (2\pi)^6\delta^3(\mathbf{k} -\mathbf{k}_2)\frac{U_\Pi(\tau_1, \tau_2, k)}{4\pi k^2}.
    \label{eq:upi}
\end{equation}
After averaging over highly oscillating modes, the GW spectrum today is
\begin{equation}
    h^2\Omega_{\rm GW}(\tau_0, k) \approx \frac{3k}{2}h^2\mathcal{T}_{GW}   \int_{\tau_*}^{\tau_{\rm fin}} \frac{d\tau_1}{\tau_1} \int_{\tau_*}^{\tau_{\rm fin}} \frac{d\tau_2}{\tau_2} U_\Pi(\tau_1, \tau_2, k)\cos k(\tau_1 - \tau_2),
\end{equation}
where $\mathcal{T}_{\rm GW}$ denotes the redshift from GW production time to today, and 
\begin{equation}
    h^2\mathcal{T}_{\rm GW} \equiv \left(\frac{a_*}{a_0}\right)^4\left(\frac{H_*}{H_0/h}\right)^2 \approx 1.6 \times 10^{-5}\left(\frac{100}{g_*}\right)^{1/3}\,.
    \label{eq:Gwup}
\end{equation}
Here $g_*$ is the number of degrees of freedom at $\tau_*$ and the Hubble rate today is $H_0 = h\times100~\mathrm{km}/\mathrm{s}/\mathrm{Mpc}$.
During the radiation-dominated era, the Hubble rate $H_*$ can be derived from $H_*^2 = 4\pi^3 Gg_*T_*^4/45$.

According to eq.~\eqref{eq:Gwup}, the key to obtaining the GW spectrum lies in the shear stress, $\tilde{\Pi}_{ij}$, which is determined by the energy-momentum tensor $T_{ij}$. In the case of a FOPT, the fluid and the scalar field contribute to $T_{ij}$. Therefore, we have
\begin{equation}
    T_{ij} = w\gamma^2v_iv_j + p\delta_{ij} + \partial_i\phi\partial_j\phi-\frac{1}{2}(\partial\phi)^2\delta_{ij},
\end{equation}
where $w$ is the enthalpy, $p$ is the pressure, and $\gamma^2 = 1/(1 - v^2)$.
However, recent studies~\cite{Hindmarsh:2013xza,Hindmarsh:2015qta,Hindmarsh:2017gnf} suggest that GWs sourced by sound waves dominate during a thermal FOPT.
Hence, to calculate the GWs from sound waves, we only need to consider the fluid part of the energy-momentum tensor which is
\begin{equation}
    T_{ij}^{\rm f} = w\gamma^2 v_iv_j + p\delta_{ij}.
\end{equation}
Numerical simulations (e.g. Hybrid simulations~\cite{Jinno:2020eqg,Jinno:2021ury,Wang:2024slx,Tian:2024ysd} or Higgsless simulations~\cite{Jinno:2022mie,Blasi:2023rqi}) can be employed to track the fluid dynamics and subsequently construct $T_{ij}^{\rm f}$, which is the essential quantity for determining the GW spectrum.  However, performing such numerical simulations is both computationally intensive and technically challenging. Therefore, in this paper, we adopt the SSM~\cite{Hindmarsh:2016lnk, Hindmarsh:2019phv, Guo:2020grp, Wang:2021dwl, Cai:2023guc, RoperPol:2023dzg, Giombi:2024kju} to compute the GWs generated by sound waves, as discussed in the following subsection.
In particular, we employ the enhanced SSM~\cite{RoperPol:2023dzg}, which offers a more reliable description in the low-frequency regime of the GW spectrum.

\subsection{The sound shell model}

The SSM is based on the fact that the fluid shells of compression and rarefaction waves surrounding expanding bubbles continue to propagate even after the collision of the bubble walls.  This model employs a set of approximations to enable a semi-analytical computation of the GWs produced by these sound waves, as outlined below:
\begin{itemize}
    \item {\it Fluid velocities are non-relativistic.}\\
    For non-relativistic fluid velocities, $\gamma^2\sim\mathcal{O}(1)$, the energy-momentum tensor can be approximated as $T_{ij}^{\rm f} \approx \bar{w}v_iv_j + p\delta_{ij}$, with $\bar{w}$ being the background enthalpy.
    Then, the UETC of the sheer stress $\tilde{\Pi}_{ij}$ is proportional to the four-point function of the velocity field, expressed as
    \begin{equation}
        \langle \tilde{\Pi}_{ij}\tilde{\Pi}_{ij} \rangle \propto \Gamma^2 \langle \tilde{v}_i\tilde{v}_j\tilde{v}_i\tilde{v}_j \rangle,
        \label{eq:pipv}
    \end{equation}
    with the adiabatic index given by $\Gamma \approx {\bar{w}}/\bar{\rho}$.
    By defining the energy fluctuation variable as $\lambda (\tau, \mathbf{x})\equiv(\rho(\tau, \mathbf{x}) - \bar{\rho})/\bar{w}$, the evolution of velocity field in momentum space can be approximately obtained by linearizing fluid equations $\partial^\mu T_{\mu\nu}^{\rm f} = 0$ as
    \begin{align}
        \tilde{\lambda}'(\tau, \mathbf{k}) - ik_i \tilde{v}^i (\tau, \mathbf{k}) &= 0, \label{eq:f1}\\
        \tilde{v}_i'(\tau, \mathbf{k}) - ik_ic_s^2\tilde{\lambda}(\tau, \mathbf{k}) &= 0.\label{eq:f2}
    \end{align}
    Here the sound speed, defined $c_s^2\equiv \bar{p}/\bar{\rho}$, is determined by the background pressure $\bar{p}$ and energy density $\bar{\rho}$. During the radiation-dominated era, $c_s^2 = 1/3$.
    
    \item {\it The fluid velocity field is Gaussian, irrotational and statistically homogeneous.}\\ 
    Under this assumption, the two-point correlation of the shear stress, expressed as the four-point correlation of the velocity field, can be further decomposed into the product of two-point functions of the velocity field:
    \begin{equation}
        \langle \tilde{v}\tilde{v}\tilde{v}\tilde{v} \rangle \sim \langle \tilde{v}\tilde{v} \rangle\langle \tilde{v}\tilde{v} \rangle.
    \end{equation}
    By defining the UETC of the velocity field $U_{v}$ as
    \begin{equation}
        \big\langle \tilde{v}(\tau_1,\mathbf{k})\tilde{v}(\tau_2, \mathbf{k}_2) \big\rangle = (2\pi)^6 \hat{k}_i\hat{k}_j\delta^3(\mathbf{k} - \mathbf{k}_2)\frac{U_{v}(\tau_1, \tau_2, k)}{2\pi k^2}\,,
    \end{equation}
    and utilizing eq.~\eqref{eq:upi} and eq.~\eqref{eq:pipv}, $U_\Pi$ can be computed as 
    \begin{equation}
        U_\Pi(\tau_1, \tau_2, k) = 2k^2\Gamma^2\int_{-1}^1dz\int_0^\infty dp \frac{p^2}{\tilde{p}^4}(1-z^2)^2U_v(\tau_1, \tau_2, p)U_v(\tau_1, \tau_2, \tilde{p}),
    \end{equation}
    where $z = \hat{k}\cdot\hat{p}$ and $\tilde{p}^2\equiv|\mathbf{k}-\mathbf{p}|^2 = p^2 + k^2 - 2pkz$.
    \item {\it The velocity field is represented as a linear superposition of self-similar fluid profiles, each associated with $N$ randomly positioned expanding bubbles nucleated at different times.}\\ 
     The solution of the velocity field from eqs.~\eqref{eq:f1} and \eqref{eq:f2} can thus be obtained from summing over fluid velocity fields generated by each bubble, denoted as  $\tilde{v}_i^{(n)}(\tau, \mathbf{k})$ for the $n$-th bubble, yielding
    \begin{equation}
        \tilde{v}(\tau, \mathbf{k}) = \sum_{s = \pm} A_s(\mathbf{k})e^{isc_sk(\tau - \tau_*)} \sim \sum_{n =1}^N \tilde{v}_i^{(n)}(\tau, \mathbf{k}).
    \end{equation}
    By utilizing the self-similar fluid profiles as initial conditions for the velocity field $\tilde{v}(\tau, \mathbf{k})$,  we obtain
    \begin{equation}
        A_\pm(\mathbf{k}) = \sum_{n = 1}^N\mathcal{A}_\pm(\chi)(T_i^n)^2 e^{i\mathbf{k}\cdot\mathbf{x}_0^n}, \quad \mathrm{with}\quad \mathcal{A}_\pm(\chi) = -\frac{i}{2}[f'(\chi) \pm ic_sl(\chi)],
    \end{equation}
    with $T_i^n$ and $\mathbf{x}_0^n$ denoting lifetime and the nucleation location of the $n$-th bubble, respectively, with $\chi\equiv kT_i^n$.
    Note that $'\equiv d/d\chi$ here. Moreover, we have 
    \begin{equation}
        f(\chi) \equiv \frac{4\pi}{\chi}\int_0^{\infty} d\xi v_{\rm ip}(\xi) \sin(\chi\xi), \quad l(\chi) \equiv \frac{4\pi}{\chi}\int_0^{\infty} d\xi \lambda_{\rm ip}(\xi) \sin(\chi\xi) \, ,
    \end{equation}
    where $v_{\rm ip}(\xi)$ and $\lambda_{\rm ip}(\xi)$ represent the velocity profile and energy fluctuation profile of a single bubble, respectively.
    Conventionally, the bag model of equation of state is employed to describe FOPTs. 
    Therefore, once $\alpha$ and $v_w$ are specified, one can obtain the above self-similar profiles with the method used in refs.~\cite{Espinosa:2010hh,Wang:2020nzm}.
    \item {\it The velocity correlator remains stationary, and bubbles are entirely annihilated once half of their volume has transitioned into the stable phase.}\\
   These assumptions enable further simplification of UETC for the velocity field,
    \begin{equation}
        U_v(\tau_1, \tau_2, k) \approx P_v(k)\cos(kc_s\tau_-),
    \end{equation}
    where $\tau_- = \tau_2 - \tau_1$ and 
    \begin{equation}
        P_v(k) = \frac{k^2}{2\pi^2\beta^6R_*^3}\int_0^\infty d\tilde{T}\nu(\tilde{T})\tilde{T}^6|\mathcal{A}(\tilde{T}k/\beta)|^2.
    \end{equation}
    Here $\beta$ is the inverse duration of phase transition, $\tilde{T} \equiv \beta T_i^n$ is the scaled bubble lifetime, and $R_*\equiv(8\pi)^{1/3}v_w/\beta$ denotes the mean bubble separation, where $v_w$ is the bubble wall velocity obtained by evaluating the friction acting on the wall~\cite{Moore:1995si,Wang:2020zlf,Jiang:2022btc}.
    The bubble lifetime, $\nu(\tilde{T})$, is derived based the above assumptions for the destruction of the bubble and some key quantities describing the dynamics of phase transition (e.g. the bubble nucleation rate, bubble wall velocity, etc.), see details in ref.~\cite{Hindmarsh:2019phv}. 
    In this work, we only consider the exponential nucleation criterion, which means $\nu(\tilde{T}) = \exp(-\tilde{T)}$.
\end{itemize}
With these approximations, we define the $R_*$-scaled spectrum of the velocity field as $\bar{P}_v = P_v/R_*$. Moreover, we introduce $\zeta_{v} \equiv \bar{P}_v/{\bar{P}_v^{\rm max}}$  to denote the normalised spectrum of the velocity field.
Using the notation of ref.~\cite{RoperPol:2023dzg}, the final GW spectrum is thus
\begin{equation}
    h^2\Omega_{\rm GW}(\tau_0, K) = 3\Gamma^2 h^2\mathcal{T}_{\rm GW} \tilde{\Omega}_{\rm GW},
    \label{eq:GWssm}
\end{equation}
where 
\begin{equation}
    \tilde{\Omega}_{\rm GW} = K^3(\bar{P}_v^{\rm max})^2\int_0^\infty dP P^2 \zeta_{v}(P) \int_{-1}^1(1 - z^2)^2 \frac{\zeta_{v}(\tilde{P})}{\tilde{P}^4}\Delta(\tau_{\rm sw},R_*, K, P, \tilde{P})dz,
\end{equation}
with
\begin{equation}
    \Delta(\tau_{\rm sw}, R_*, K, P, \tilde{P}) \equiv \int_{\tau_*}^{\tau_{\rm fin}}\frac{d\tau_1}{\tau_1}\int_{\tau_*}^{\tau_{\rm fin}}\frac{d\tau_2}{\tau_2}\cos\left(Pc_s\frac{\mathcal{H}_*\tau_-}{\mathcal{H}_*R_*}\right)\cos\left(\tilde{P}c_s\frac{\mathcal{H}_*\tau_-}{\mathcal{H}_*R_*}\right)\cos\left(K\frac{\mathcal{H}_*\tau_-}{\mathcal{H}_*R_*}\right).
    \label{eq:kernel}
\end{equation}
Here the lifetime of sound waves is defined as $\tau_{\rm sw} \equiv \tau_{\rm fin} - \tau_*$, and  $c_s$ denotes the sound speed.
To facilitate numerical integrations, we set $K = kR_*$, $P = pR_*$, and $\tilde{P} = \tilde{p}R_*$.
Note that the main difference between the original SSM~\cite{Hindmarsh:2016lnk,Hindmarsh:2019phv} and the enhanced SSM~\cite{RoperPol:2023dzg} lies in their treatment of the kernel given in eq.~\eqref{eq:kernel}, which leads to different low-frequency scaling behaviour. 
Specifically, the original SSM approximates the kernel~\eqref{eq:kernel} as a delta function, while the enhanced SSM precisely incorporates it into the calculation. 

With all these ingredients, we can finally obtain the GW spectrum today by computing the red-shifted frequencies as
\begin{equation}
    f = 2.6\times10^{-6}\mathrm{Hz}\frac{K}{\mathcal{H}_*R_*}\left(\frac{T_*}{100~\mathrm{GeV}}\right)\left(\frac{g_*}{100}\right)^{1/6}.
\end{equation}

However, the high-dimensional numerical integrations involved in Eq.~\eqref{eq:GWssm} cannot be performed efficiently, and the oscillatory nature of the integration kernel further complicates and slows down the entire computation.
We thus seek to construct an emulator based on NN for the SSM in the next section.

\section{The NN architecture and the training strategy}
\label{sec:NN}


\begin{figure}[t!]
    \centering
    \includegraphics[width=0.9\linewidth]{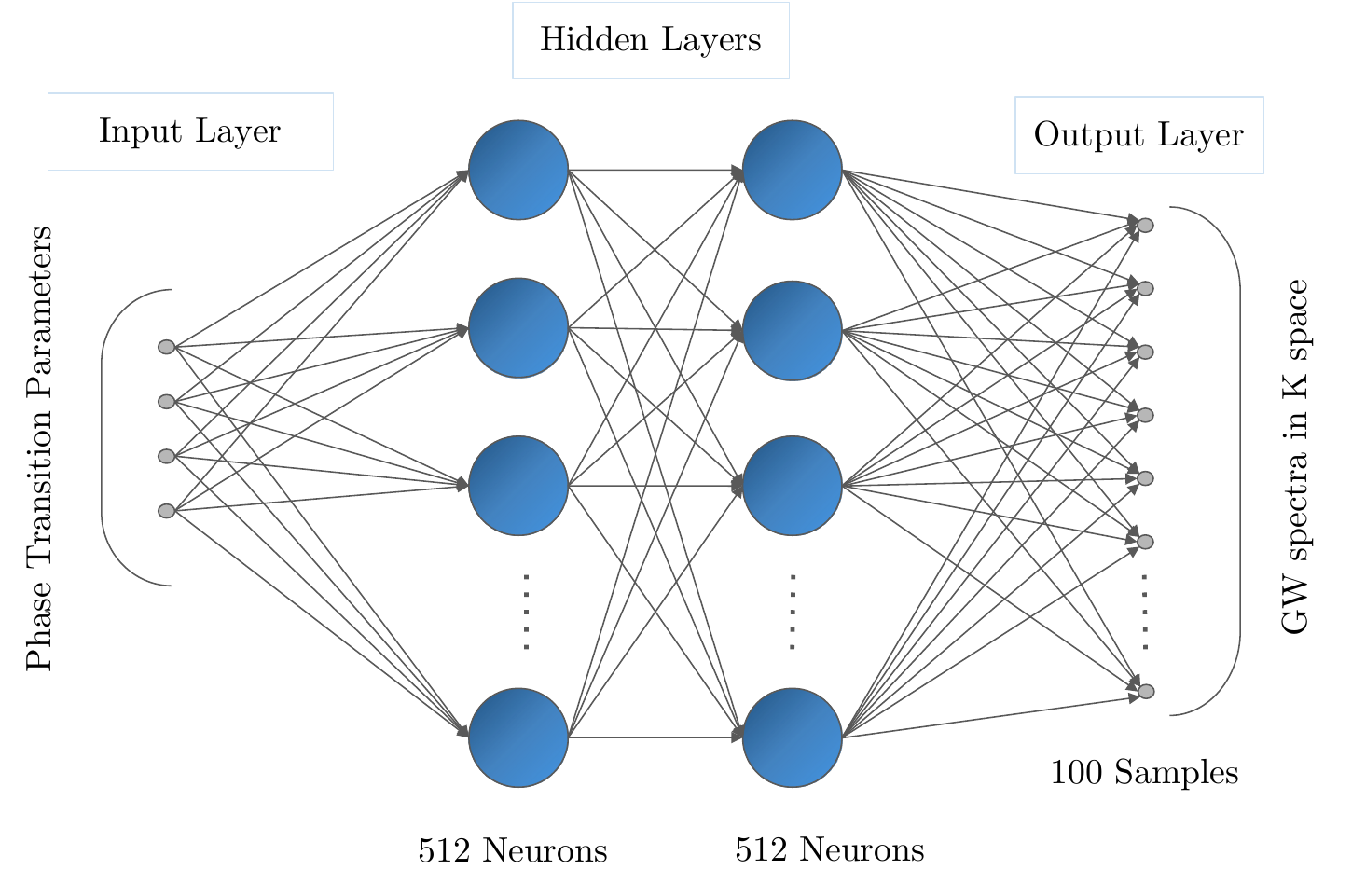}
    \caption{The schematic of the architecture of the NN used by the emulator. The input layer accepts phase transition parameters $v_w,\, \alpha\,,\mathcal{H}_* \tau_{\rm sw}\,,\mathcal{H}_* R_*$; the output layer generates GW spectrum $\tilde{\Omega}(K)$ at $100$ points uniformly distributed in the log-space of $K$.}
    \label{fig:NN_arch}
\end{figure}

The architecture of the NN used in our emulator is shown in Fig.~\ref{fig:NN_arch}.  It comprises multiple layers of fully connected neurons.  The input layer accepts $4$ phase transition parameters $v_w$, $\alpha$, $\mathcal{H}_* \tau_{\rm sw}$, $\mathcal{H}_* R_*$.  The output layer generates the complete spectrum of $\tilde{\Omega}_{\mathrm{GW}}(K)$, sampled at $100$ logarithmic evenly spaced points in the dimensionless $K$--space, ranging from $1\times10^{-3}$ to $3\times10^2$.  Each of the two hidden layers contains 512 neurons.
The projection corresponding to this NN is strictly linear until an activation function is introduced.  We adopt the Rectified Linear Unit (ReLU) as the activation function.  Both the NN architecture and the activation function are selected based on our experimental results, which demonstrate their superior accuracy compared to other commonly adopted alternatives.  It is worth noting that the NN does not take the phase transition temperature $T_*$ or the number of degrees of freedom $g_*$ as inputs, as the resulting gravitational wave spectrum today depends linearly on these parameters, as predicted by the enhanced SSM.

We utilised the enhanced SSM code to generate the datasets used for training and testing.  The training set consists of $400,000$ data sets, with phase transition parameters randomly sampled uniformly or log-uniformly.  The range and sampling scheme for these parameters are detailed in Table~\ref{tab:params}. Note that while the ranges of the phase transition parameters are chosen to cover a reasonably large parameter space that is practical, the phase transition strength is set to $\alpha < 1/3$.  This limitation arises due to the existence of a lower bound on the wall velocity, as predicted by the bag model of the equation of state for $\alpha > 1/3$~\cite{Espinosa:2010hh}, which introduces additional ambiguities to the NN training and the Bayesian inference.  Moreover, as the strength of the phase transitions increases, the non-relativistic approximation for the fluid velocity may become unreliable.  To account for these complexities, we impose this upper limit for simplicity, with the intention of expanding the data range in future work.
Meanwhile, the lifetime of sound wave $\mathcal{H}_*\tau_{\rm sw}$ is treated as a free parameter in our work, while it can be estimated based the mean bubble separation $\mathcal{H}_*R_*$, see details in ref.~\cite{Hindmarsh:2019phv}.
It is also important to note that the parameter ranges defined for the training set also establish the recommended bounds for the input parameters of the neural network, as well as the priors for the subsequent Bayesian analysis, which are distributed either uniformly or log-uniformly within these intervals.  Furthermore, 20\% of the training dataset is reserved for validation purposes rather than used directly for training. 

\begin{table}[t!]
\centering
\begin{tabular}{|c|c|c|}
    \hline
     Parameter& Sampling range & Sampling mode\\
     \hline
     $v_w $  &  $[0.01 , 0.99]$ & uniform \\
     $\alpha $  &  $[0.001 , 0.33]$ & uniform \\ 
     $\mathcal{H}_*\tau_{\rm sw} $  &  $[0.0001 , 1]$ & log-uniform \\
     $\mathcal{H}_* R_* $  &  $[0.0001 , 1]$ & log-uniform \\
     \hline
\end{tabular}
\caption{Ranges and random sampling mode of the phase transition parameters for training.}
\label{tab:params}
\end{table}

Our training process uses a mean squared error loss function and the Adam optimizer~\cite{Kingma2014AdamAM}, initialised with a learning rate of $0.01$ and a batch size of $1024$.  The entire training procedure lasts for 5 rounds, each consisting of hundreds to thousands of training epochs, with the early stopping triggered when validation loss increases over consecutive epochs.  Each subsequent training round resumes with the learning rate reduced by half, and the batch size doubled.  It is worth noting that, while the semi-analytical SSM model generates training and testing datasets in double precision, the neural network operates in single precision, which is sufficient given the maximum precision achievable by the emulator.  Although generating the training and testing datasets requires approximately $200,000$ CPU hours, the complete training process only takes about six hours on a single NVIDIA RTX 4090 GPU device.

\begin{figure}[t!]
    \centering
    \includegraphics[width=0.7\linewidth]{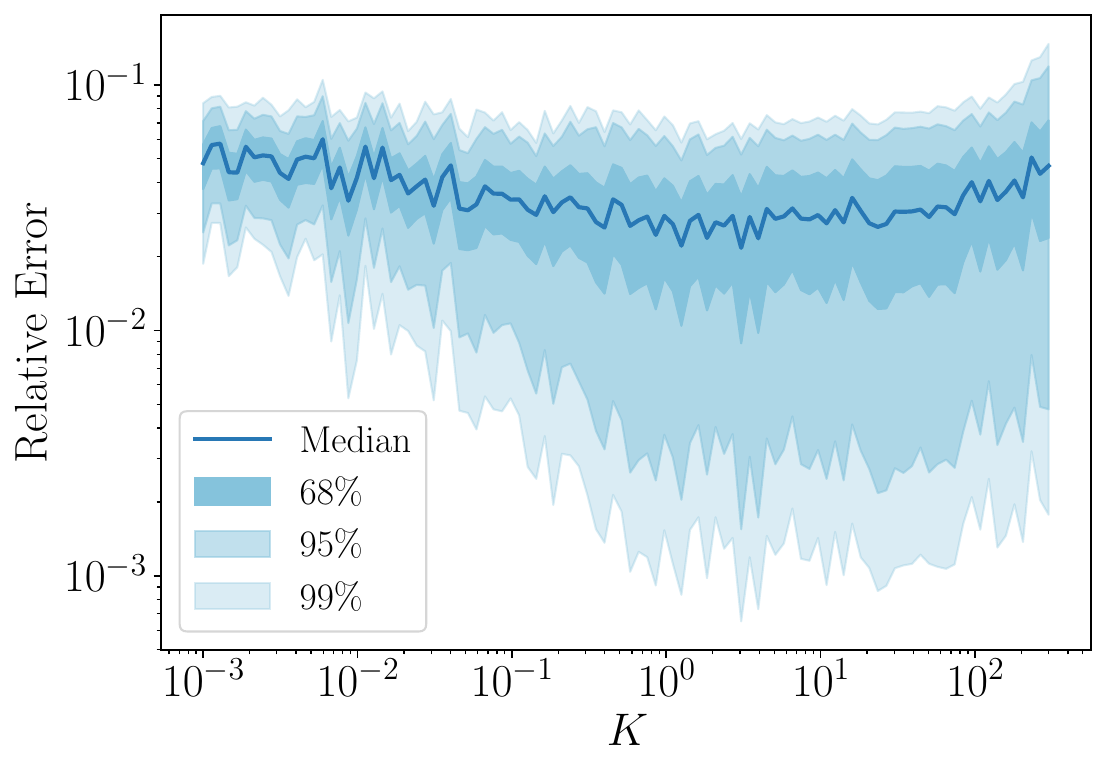}
    \caption{Relative error, defined  $ |(\Omega_{\mathrm{GW}}^{\mathrm{NN}} - \Omega_{\mathrm{GW}}^{\mathrm{SSM}})/ \Omega_{\mathrm{GW}}^{\mathrm{SSM}}|$ for the NN predictions on a test set of size $20,000$. The shaded regions indicate the 68th, 95th, and 99th percentiles of the relative error within the test set.}
    \label{fig:NN_test}
\end{figure}

After completion of the entire training process, the performance of the trained neural network is evaluated on a test set composed of additional $20,000$ independently and randomly sampled parameter sets that are not included in the training dataset.
The relative errors between the NN predictions and the semi-analytical results are summarised in Fig.~\ref{fig:NN_test}. Evidently, the NN demonstrates a high level of accuracy in approximating the output of the enhanced SSM model, achieving a median error of less than $5\%$, with $99\%$ of the test cases exhibiting relative errors below $10\%$.  In terms of computational efficiency, the NN requires no more than $100~\rm ms$ per calculation, even when executed on a CPU.  In comparison, the semi-analytical SSM model, implemented by a highly optimised Python-wrapped C code, still costs 5--30 minutes per run.  Thus, the emulator provides several orders of magnitude in computational speed-up, significantly alleviating heavy computational demands, such as extensive parameter space scanning or Bayesian analysis.

\section{Inferring phase transition parameters from mock LISA data}
\label{sec:LISA}

In this section, we demonstrate the potential applications of our emulator, employing it to directly reconstruct the phase transition parameters from mock LISA data through Bayesian inference, without relying on any empirical models.

\subsection{Estimating LISA noise and generating mock data}

The Laser Interferometer Space Antenna (LISA) mission, scheduled for launch in the mid-2030s, will comprise three satellites arranged in a nearly equilateral triangular formation, with arm lengths of 2.5 million kilometres.  We follow the approach depicted in \cite{Caprini:2019pxz, Giese:2021dnw, Flauger:2020qyi} to construct LISA mock data. The noise is estimated based on a simplified model, which asserts that the total Time-Delay Interferometry (TDI) noise power spectrum density (PSD) of LISA contains two dominant contributions: the optical metrology system noise PSD $P_{\rm oms}(f,P)$, and the mass acceleration noise PSD $P_{\mathrm acc}(f,A)$.  These noise components are defined as:

\begin{align}
\label{eq:PP}
    P_{\rm oms} (f,P) &= P^2 \frac{\rm pm^2}{\rm Hz} \left(1 + \left(\frac{2 \rm \, mHz}{f} \right)^4 \right) \left(\frac{2\pi f}{c} \right)^2,  \\
	P_{\rm acc} (f,A)&= A^2 \frac{\rm fm^2}{\rm s^4 Hz} \left(1 + \left(\frac{0.4 \rm \, mHz}{f} \right)^2 \right)  \\
& \times \left(1 + \left(\frac{f}{8 \rm \, mHz} \right)^4 \right) \left(\frac{1}{2\pi f} \right)^4 \left(\frac{2\pi f}{c} \right)^2.
\end{align}
They dominate at higher and lower frequencies, respectively, with $P \approx 15$, $A \approx 3$ as noise parameters.  Based on this model, under the  equilateral triangle assumptions of the three satellite and additional isotropic and stationary assumptions on $P_{\rm oms}(f,P)$ and $P_{\mathrm acc}$ \footnote{We refer to \cite{Flauger:2020qyi} for  further details on noise estimations}, the total noise in the measured energy fraction of GWs can be estimated as:
\begin{align}
    \Omega_{\rm noise}(f) = \frac{4 \pi^2}{3 H_0^2} f^3 \frac{\frac{10}{3}\left(1+ 0.6 \left(\frac{2 \pi f L}{c} \right)^2 \right)\left(P_{\rm oms} (f,P) + \left( 3 + \cos{\frac{4\pi f L}{c}}\right)P_{\rm acc}(f,A) \right)}{(2 \pi fL/c)^2}\,,
\end{align}
where $H_0$ is the Hubble parameter and $L\approx  2.5 \times 10^6 ~ \mathrm{km} $ is the arm length. 

The simulated signals $\Omega_{\rm GW}(f)$, used as test cases, are generated based on the enhanced SSM based on two representative sets of benchmark parameters. The benchmarks, $\mathrm{BP}_1$ and $\mathrm{BP}_2$, are listed in Table~\ref{tab:BPs}. They represent relatively strong phase transitions that occur during the electroweak epoch, resulting in a power-law-like signal and a comparatively flatter signal, respectively (see the blue curves in Fig.~\ref{fig:dat_shape}).  The differences in their spectral shapes are mainly attributed to their distinct bubble wall velocities $v_w$. It is important to note that $\rm{BP}_1$ corresponds to the deflagration case with relatively strong phase transition strength. In this instance, due to the loss of kinetic energy, the GW content could be significantly suppressed compared to the predictions of the SSM. This discrepancy was first identified in \cite{Cutting:2019zws} through hydrodynamical simulations and can be addressed by introducing a suppression factor into the SSM \cite{Gowling:2021gcy}. For now, this suppression has not been considered in the enhanced SSM code or the DeepSSM emulator. Therefore, caution should be made when employing this emulator, a suppression factor (such as the one proposed in  \cite{Gowling:2021gcy}) may need to be manually introduced when exploring the parameter region with significant suppression. We plan to update both the enhanced SSM model and the emulator to accommodate this suppression effect in future work.

We generate the injected mock data using the GW spectra corresponding to these two benchmark points, calculated with the enhanced SSM code rather than the emulator.  
Any foreground signals, such as those introduced by 
galactic binaries, are approximately ignored in this work for simplicity. Techniques such as yearly modulation have proven highly effective in mitigating such foreground, particularly for power-law signals~\cite{Adams:2013qma}. However, caution should be made when addressing peaked signals associated with phase transitions. As noted by~\cite{Hindmarsh:2024tt}, the broken power-law signal resulting from a FOPT can be approximately recovered for  $\Omega_{\rm GW, peak} \geq 10^{-11}$ and $f_{\rm peak} \geq 2\times10^{-3}\,\rm Hz$ from galactic binary foreground signals after yearly modulation, a threshold closely aligns with our benchmark points. Given the complex GW spectral shapes predicted by our model, when using our framework on signals contaminated by foreground noise, more careful studies are necessary to accurately extract phase transition parameters.

\begin{table}[t!]
\centering
\begin{tabular}{|c|c|c|c|c|c|}
    \hline
      & $v_w$ & $\alpha$ & $\mathcal{H}_* \tau_{\rm sw}$ & $\mathcal{H}_*R_*$ & $T_*~[\mathrm{GeV}]$ \\
      \hline
      $\mathrm{BP}_1$ & 0.15 & 0.3 & 0.8 & 0.002 & 120\\
      $\mathrm{BP}_2$ & 0.6 & 0.3 & 0.8 & 0.008 & 100\\
      \hline
\end{tabular}
\caption{Phase transition parameters of our benchmark points.}
\label{tab:BPs}
\end{table}

After building the signal and noise model, our mock data is produced in frequency bins spanning between $f=[3\times10^{-5}, 0.1]~\rm Hz$, with $\Delta f = 10^{-6}~\mathrm{Hz}$, which corresponds to an observational segment of $T_{\mathrm seg}\sim 11 $ days.  Taking into account $75\%$ observational efficiency, it corresponds to $N_{\mathrm{c}} = 94$ data points in each frequency bin.  For each frequency bin $f_{i}$, we generate $N_{\mathrm{c}}$ signal and noise data points, following the distribution:
\begin{align}
    	&S_{i} = \left \rvert \frac{G_{i1}\left(0,\sqrt{h^2 \Omega_{\rm GW}(f_i)}\right) + iG_{i2}\left(0,\sqrt{h^2 \Omega_{\rm GW}(f_i)} \right)}{\sqrt 2} \right\rvert^2 , \\
	& N_{i} = \left \rvert \frac{G_{i3}\left(0,\sqrt{h^2 \Omega_{\rm noise}(f_i)}\right) + iG_{i4}\left(0,\sqrt{h^2 \Omega_{\rm noise}(f_i)} \right)}{\sqrt 2} \right\rvert^2,
\end{align}
where $i$ is the index of the frequency bins and $G_{ik}$ represents a real number randomly drawn from a normal distribution.  We then set the signal $D_{i}$ to be
\begin{align}
    D_{i} = S_{i} + N_{i}\,,
\end{align}
and finally take the average of total $N_{\mathrm{c}}$ data points in each frequency bin, denoted as $\bar{D}_{i}$, as our data.  Note that the data set generated by this scheme does not strictly follow a Gaussian distribution, although the deviations are minor due to the central limit theorem.  We will take this into account when formulating our likelihood function.  
In addition, unlike the approach taken in \cite{Caprini:2019pxz, Giese:2021dnw, Flauger:2020qyi}, we do not perform additional coarse-graining to the datasets to reduce computational demands.  This is because our NN based emulator with the HMC Bayesian framework performs efficiently, as demonstrated in the following subsection.

\subsection{Bayesian inference}

\begin{figure}[t!]
    \centering
    \includegraphics[width=0.49\linewidth]{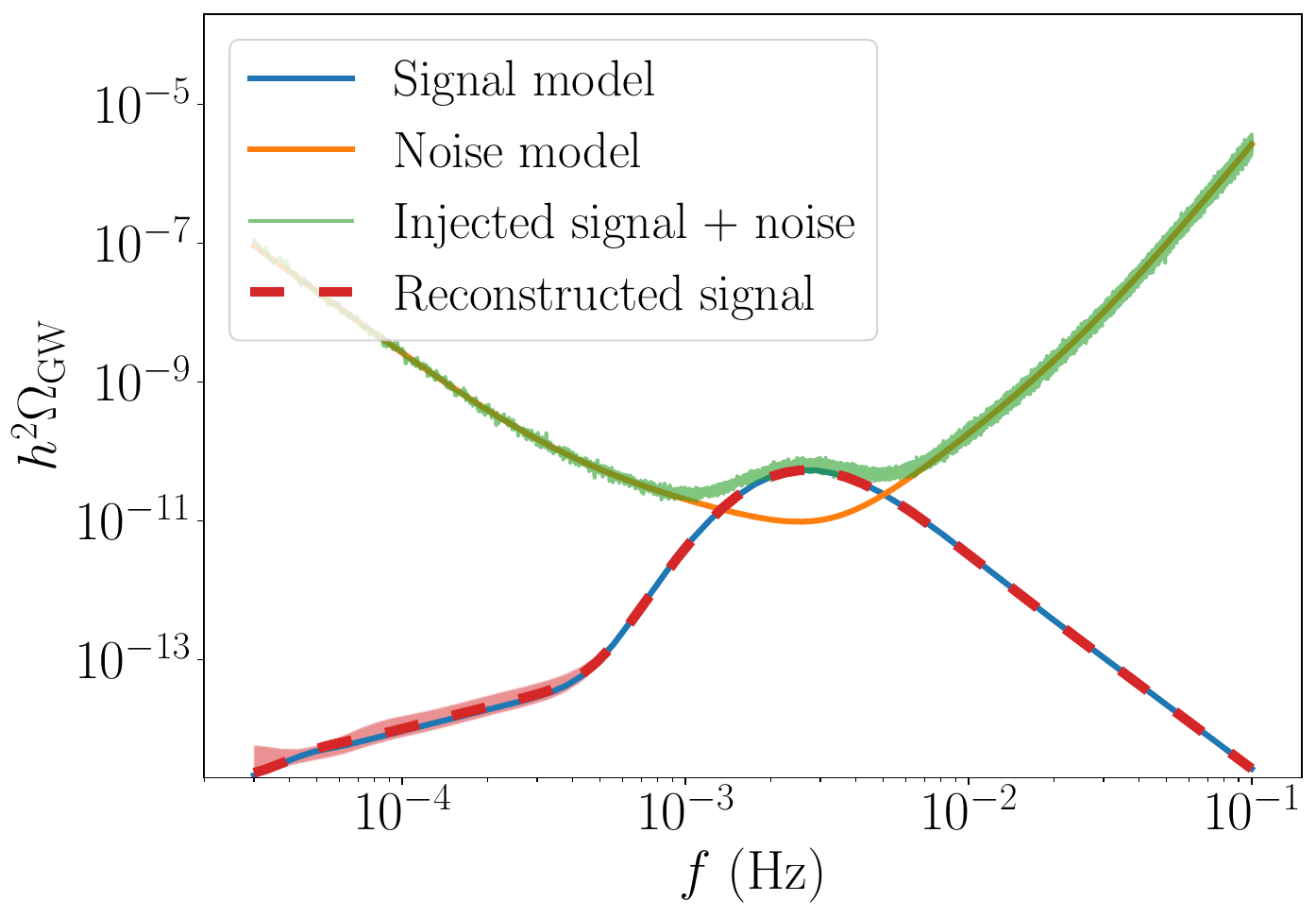}
    \includegraphics[width=0.49\linewidth]{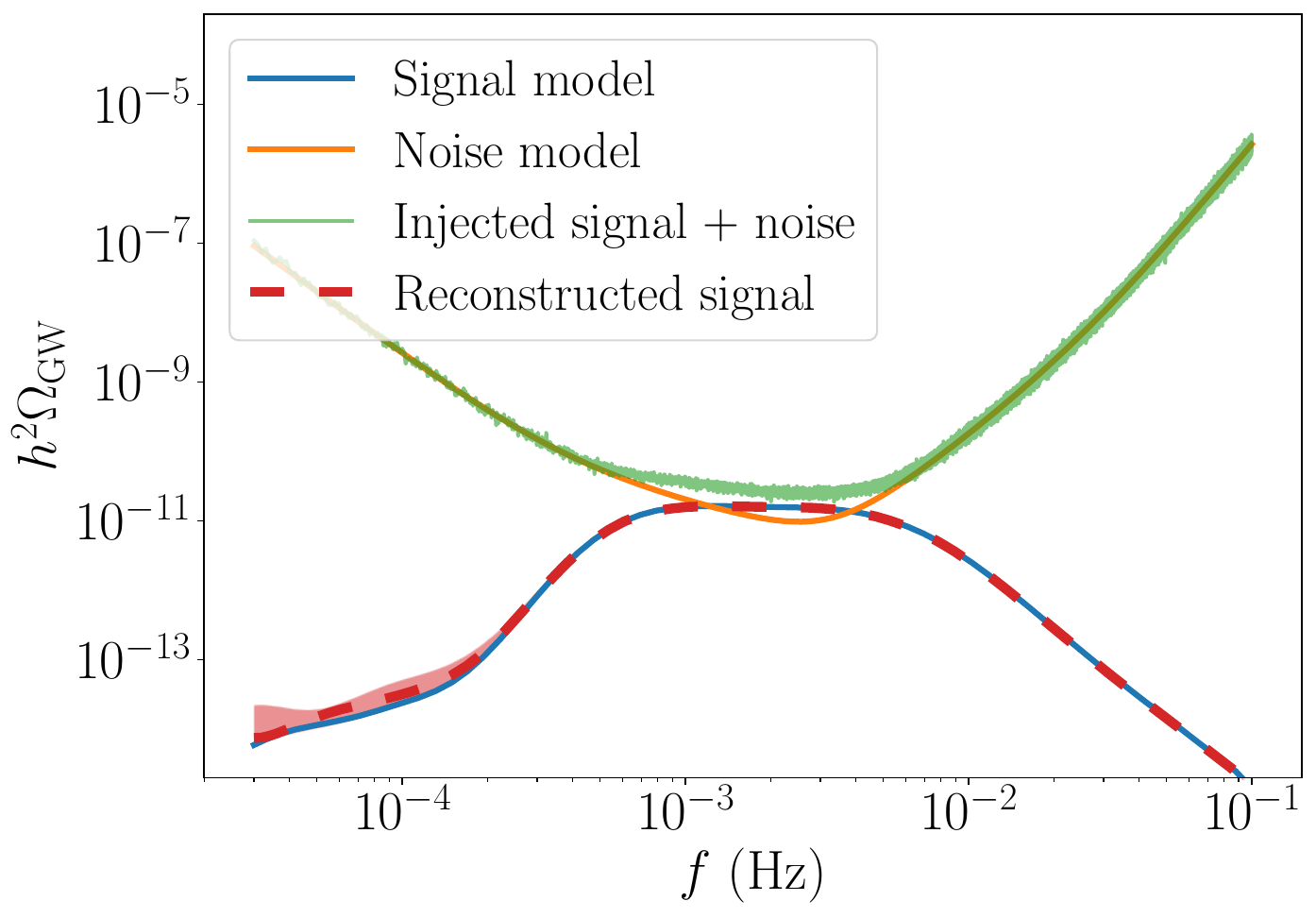}
    \caption{The signal and LISA noise model (blue and orange smooth curves) alongside the injected mock data ($\bar{D}_i$, wiggly green curves) and the reconstructed GW spectra. The left and right panels correspond to the benchmark points $\mathrm{BP}_1$ and $\mathrm{BP}_2$, respectively. The red shaded  regions represent the $1\sigma$ reconstruction uncertainties of the  spectra.}
    \label{fig:dat_shape}
\end{figure}

We now employ the Markov Chain Monte Carlo (MCMC) method to reconstruct the model parameters from the injected mock LISA data.  They are
\begin{align}
    v_w,\;\alpha,\;\mathcal{H}_*\tau_{\rm sw},\;\mathcal{H}_*R_*,\;T_*,\;P,\;A\,,
\end{align}
where the first five parameters are the phase transition parameters and the rest two are noise parameters.  The fiducial values for the phase transition parameters are listed in Table~\ref{tab:BPs} with $g_*=100$, and the fiducial noise parameters are set to be $P=15$ and $A=3$ for both cases.

To reconstruct these model parameters, first we must define a likelihood function that accounts for the non-Gaussianities in the generated mock data.  Following \cite{Flauger:2020qyi}, which adopts techniques from CMB analysis \cite{WMAP:2003pyh}, we utilize the logarithmic likelihood function as 
\begin{align}
\label{eq:lihd}
\log \mathcal{L}_\mathrm{G+LN}\equiv \frac{1}{3} \log \mathcal{L}_\mathrm{G}+\frac{2}{3}\log \mathcal{L}_\mathrm{LN}\, ,
\end{align}
where 
\begin{align}
  &  \log\mathcal{L}_\mathrm{G}=-\frac{N_c}{2}\sum_i \left(\frac{\bar{D_i}-h^2\Omega_{\mathrm{noise}}(f_i)-h^2\Omega_{\mathrm{GW}}(f_i)}{ h^2\Omega_{\mathrm{noise}}(f_i) + h^2\Omega_{\mathrm{GW}}(f_i) }\right)^2\, , \\
    &\log\mathcal{L}_\mathrm{LN}=-\frac{N_c}{2}\sum_i \left(\log \frac{h^2\Omega_{\mathrm{noise}}(f_i) + h^2\Omega_{\mathrm{GW}}(f_i)}{\bar D_i}\right)^2\, .
\end{align}
They account for the Gaussian and marginally non-Gaussian contributions to the likelihood, respectively.  As suggested by \cite{Giese:2021dnw}, compared to the conventional $\chi^2$ likelihood, this modified likelihood function generally outperforms the standard Gaussian likelihood in providing unbiased estimates of model and noise parameters for most analyses.

\begin{figure}[t!]
    \centering
    \includegraphics[width=1\linewidth]{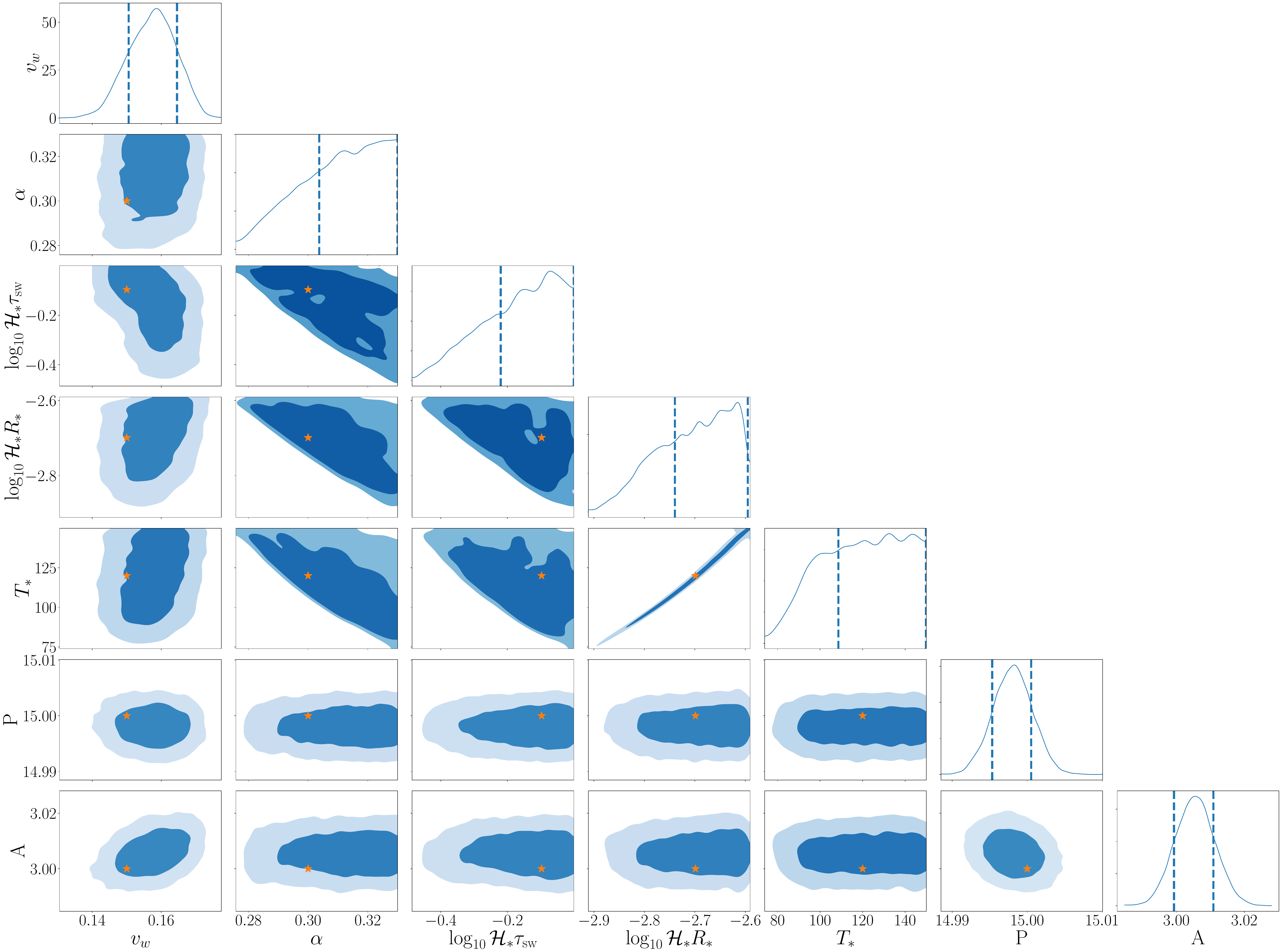}
    \caption{Contours and marginalized posterior distributions of phase transition and noise parameters of $\mathrm{BP}_1$.  Stars indicate the fiducial values of the model parameters.  Dashed vertical lines in the marginalized posteriors represent $1\sigma$ bounds.  The dark and light blue shaded regions denote the $1\sigma$ and $2\sigma$ confidence intervals, respectively.}
    \label{fig:ct1}
\end{figure}

\begin{figure}[t!]
    \centering
    \includegraphics[width=1\linewidth]{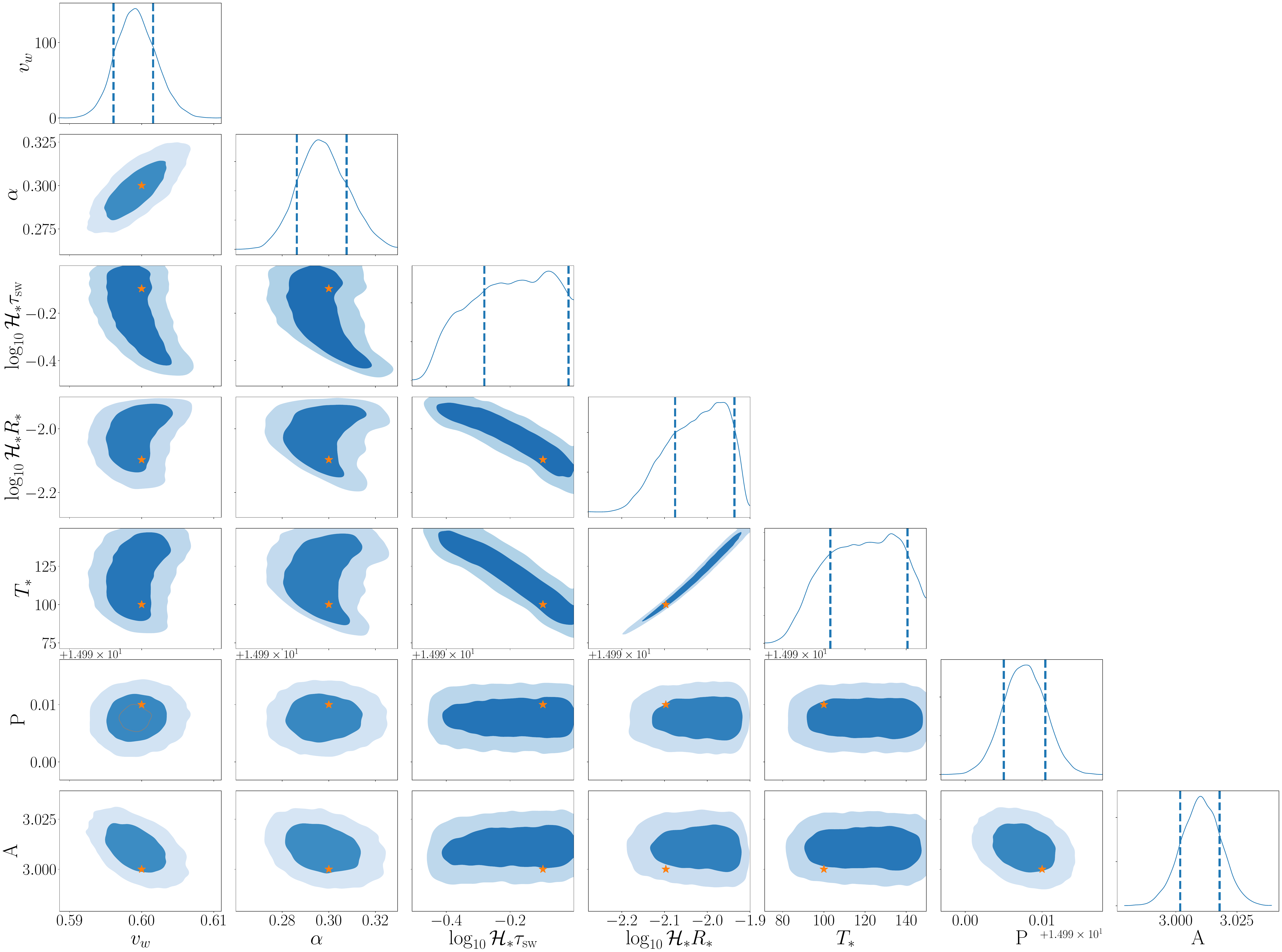}
     \caption{Contours and marginalized posterior distributions of phase transition and noise parameters of $\mathrm{BP}_2$.  Stars indicate the fiducial values of the model parameters.  Dashed vertical lines in the marginalized posteriors represent $1\sigma$ bounds.  The dark and light blue shaded regions denote the $1\sigma$ and $2\sigma$ confidence intervals, respectively.}
    \label{fig:ct2}
\end{figure}

We performed a Bayesian analysis to reconstruct the fiducial parameter values and explored their degeneracies.  In the Bayesian framework employed in this work, instead of using traditional MCMC samplers, which suffer from the inefficiency caused by the random walk behaviour of the Metropolis algorithm, we adopt the No-U-Turn Sampler (NUTS~\cite{hoffman2014no}), an advanced extension of the HMC~\cite{1987PhLB..195..216D} sampler.  Based on the basic principle of Hamiltonian mechanics, these modern samplers have demonstrated substantial performance improvements over traditional approaches.  However, applications of these samplers are often challenging due to their requirements for gradient information of the likelihood function.  Due to the automatic differentiation capabilities of NNs via backpropagation~\cite{rumelhart1986learning}, NNs intrinsically provide gradient information, making them particularly well-suited for gradient-based inference schemes such as the NUTS.

The results for the reconstruction of the GW spectrum are shown in Fig.~\ref{fig:dat_shape}, where the signal and noise model of $\mathrm{BP}_1$ and $\mathrm{BP}_2$ are represented by smooth curves, and the injected mock data ($\bar{D}_i$) are displayed as green wiggly curves, reflecting the fluctuations in the realistic data.  It is evident that, regardless of the shape of the injected spectrum, with LISA noise, the GW spectra produced by both of our benchmark models can be effectively reconstructed using \texttt{DeepSSM}.
The reconstruction accuracy shows only a slight decline in the low-frequency regime, as indicated by the red-shaded uncertainty region.

The contour plots of the posteriors for the phase transition and noise parameters are shown in Fig.~\ref{fig:ct1}  and Fig.~\ref{fig:ct2}, for $\mathrm{BP}_1$ and $\mathrm{BP}_2$, respectively.  Notably, the noise parameters $A$ and $P$ are constrained very tightly by the mock data, while the phase transition parameters exhibit varying levels of uncertainty.
For most parameters, the fiducial values are successfully reconstructed within $1\sigma$ uncertainty.  However, there are minor biases for the parameters $\mathcal{H}_*R_*$ and $T_*$ in $\rm BP_2$, which are two highly degenerate parameters and have fiducial values located slightly outside the $1\sigma$ regions.  This could be attributed to the low signal-to-noise ratio in the frequency ranges where the most constraining power comes from. 

Furthermore, by comparing the contour plots for these two benchmark cases, discrepancies are observed in the degeneracy patterns, such as $\alpha$ vs. $v_w$ , $T_*$ vs. $\alpha$ or $\mathcal{H}_* R_*$ vs $\alpha$, which may arise from the distinct spectral shapes of these two test cases.  In addition, the constraints on parameters such as $\mathcal{H}_*\tau_{\rm sw}$, $\mathcal{H}_* R_*$ or $T_*$, are weaker compared to those on $v_w$ and $\alpha$ in both cases, indicating that the constraining capability of the LISA observatory might be insufficient to fully unravel the details of cosmological first-order phase transitions.  We anticipate significant improvements from various next-generation gravitational wave observatories.

Additionally, studies such as~\cite{Gowling:2022pzb} report a bimodal posterior for a benchmark point with $v_{w}=0.55$. However, a similar bimodal posterior is not observed in either of our benchmark points. This discrepancy may be attributed to differences in our benchmark phase transition parameters  or to the difference models employed (an enhanced SSM-based emulator vs. an  model combining SSM and an empirical template). Furthermore, multimodal posteriors are known to pose significant challenges for MCMC algorithms, as these methods often become trapped in isolated modes. Advanced techniques, such as multimodal nested sampling \cite{Feroz:2007kg}, may prove more effective in addressing such scenarios. We defer a comprehensive exploration of the parameter space and an investigation of potential multimodal degeneracies to future work.

The posteriors and contours presented in this section are based on the HMC sampler with a single chain. The entire sampling process typically requires 12,000 points to complete and takes less than 3 hours. During the MCMC sampling, the exceptional efficiency of the emulator allows the GW signal spectrum to be computed instantaneously, resulting in time costs being only attributed to likelihood computation and sampling processes. This large number of sampling point will make semi-analytical model such as the enhanced SSM impractical to be used in the MCMC, as each estimation typically needs 10 minuets to accomplished. Compared to Ref.~\cite{Gowling:2022pzb}, our framework demonstrates greater efficiency, likely attributable to the emulator's high performance, the use of advanced samplers like HMC, and more cutting-edge hardware.

\section{Conclusions}
\label{sec:conc}

We introduce \texttt{DeepSSM}, an NN-based emulator developed to accelerate the computation of the SGWB spectrum generated by sound waves from first-order phase transitions.  Trained on the output generated by an enhanced SSM, \texttt{DeepSSM} is able to create an accurate nonlinear mapping between critical phase transition parameters and the resulting GW spectra.
Compared to the traditional semi-analytical approach, \texttt{DeepSSM} achieves acceleration by several orders of magnitude while incurring only a marginal reduction in accuracy, thereby enabling Bayesian analysis to be performed directly without relying on empirical models such as broken power-law relations.  Moreover, the fully differentiable nature of NNs makes them inherently compatible with advanced Monte Carlo samplers, such as NUTS or HMC, allowing for significantly more efficient MCMC sampling.  These two key advantages make \texttt{DeepSSM} highly effective and convenient for parameter estimation using current and future SGWB data, eliminating the need for empirical modelling. 

We demonstrate this capability by applying the emulator to mock LISA data, successfully reconstructing both the underlying phase transition parameters and the injected gravitational wave spectra via a NUTS sampler.  The results clearly highlight the degeneracies between phase transition parameters and noise parameters in the context of LISA, comprehensively demonstrating the sensitivity of LISA to GW spectra arising from FOPTs.  Importantly, the entire inference pipeline is generic, lightweight, and adaptable to other present or upcoming gravitational wave observatories, with the ability to run even on standard consumer laptops. 

Note that the lifetime of sound waves $\tau_{\rm sw}$ is treated as a free parameter in our analysis. In principle, it should be closely related to other phase transition parameters (i.e., $\alpha$, $R_*$ and $v_w$). For example, $\tau_{\rm sw}$ can be approximated by the eddy turnover time, defined as the time before the fluid enters the non-linear regime $\tau_{\rm nl} \sim R_* /\sqrt{\Omega_K}$ \cite{Caprini:2019egz,Hindmarsh:2017gnf}, 
where $\Omega_K$ is the kinetic energy fraction and can be calculated as shown in Refs.~\cite{Espinosa:2010hh,Giese:2020rtr,Giese:2020znk,Wang:2020nzm,Wang:2023jto}. However, to maintain generality, we choose to consider $\tau_{\rm sw}$ as a free parameter when implementing the enhanced SSM, the emulator and the Bayesian inference. Thanks to the exceptional efficiency of our emulator, this approach may facilitate research on various choices of $\tau_{\rm sw}$. For instance, researchers are able to fix $\tau_{\rm sw}$ by incorporating their own estimates of $\tau_{\rm sw}$ when adopting our framework within their Bayesian pipelines.

Apart from its application in Bayesian analysis, the emulator may have broader applications in astroparticle physics, as outlined below:
\begin{itemize}
    \item \emph{Exploration of parameters spaces in BSMs} --- Given that these phase transition parameters encapsulate critical features of BSM physics associated with FOPTs, the emulator effectively serves as a bridge connecting BSM physics to GW observables.  By substantially reducing the computational costs involved in the exploration of parameter spaces, it significantly facilitates the study of various BSMs.
    \item \emph{Foreground Cleaning} --- In cases where only weak FOPTs occur during the radiation era, the resulting signals may exhibit strengths comparable to or weaker than those from compact binaries, effectively acting as a foreground to astrophysical signals.  Thus, extracting astrophysical sources requires efficient foreground cleaning, which could benefit from the highly precise and computationally efficient foreground templates generated by \texttt{DeepSSM}.
\end{itemize}

There are several ways to improve this emulator.  First, the parameter ranges are currently limited by the training set, although there have been discussions on the extrapolation ability of the neural network for parameter ranges outside the training set.  To achieve enough accuracy, the safest way is still expanding the coverage of the training set. 
Note that extending the phase transition strength $\alpha$ to $\alpha > 1/3$ would require a proper strategy to organise the training set for the NN, and to perform Bayesian analysis, since there would be a lower bound for the wall velocity for $\alpha > 1/3$. We thus leave it for further study.  
Second, although challenging, training the emulator on the data produced by lattice simulations can provide a better estimation of the GW spectra.  Sophisticated NN training strategies can be useful in addressing these challenges.  For example, advanced machine learning techniques such as transfer learning \cite{pan2009survey} can be applied in this scenario by creating a base model using a semi-analytical approach and fine-tuning it with the data given by lattice simulations.  We reserve explorations of such topics for future study. 

\acknowledgments
We thank Ran Ding and Xiao-Xiao Kou for their insightful comments on this work. C.T. is supported by the National Natural Science Foundation of China (Grants No.~12405048) and the Natural Science Foundation of Anhui Province (Grants No.~2308085QA34).
X.W. and C.B. are supported by Australian Research Council grants DP210101636, DP220100643 and LE21010001.
The NN architecture and training framework used in this work are implemented using \texttt{Flax}~\cite{flax2020github}, a highly flexible NN library based on the high-computing performance framework \texttt{Jax}~\cite{jax2018github}. The Bayesian framework, based on the NUTS, is developed using \texttt{NumPyro}~\cite{phan2019composable,bingham2019pyro}, which is a \texttt{Jax} based probabilistic programming library. The contour visualization plots were created by the \texttt{ArviZ} library~\cite{Kumar2019}.


\bibliographystyle{JHEP}
\bibliography{biblio.bib}

\providecommand{\href}[2]{#2}\begingroup\raggedright\begin{thebibliography}{10}

\bibitem{NANOGrav:2023gor}
{\scshape NANOGrav} collaboration, \emph{{The NANOGrav 15 yr Data Set: Evidence for a Gravitational-wave Background}}, \href{https://doi.org/10.3847/2041-8213/acdac6}{\emph{Astrophys. J. Lett.} {\bfseries 951} (2023) L8} [\href{https://arxiv.org/abs/2306.16213}{{\ttfamily 2306.16213}}].

\bibitem{Xu:2023wog}
H.~Xu et~al., \emph{{Searching for the Nano-Hertz Stochastic Gravitational Wave Background with the Chinese Pulsar Timing Array Data Release I}}, \href{https://doi.org/10.1088/1674-4527/acdfa5}{\emph{Res. Astron. Astrophys.} {\bfseries 23} (2023) 075024} [\href{https://arxiv.org/abs/2306.16216}{{\ttfamily 2306.16216}}].

\bibitem{EPTA:2023fyk}
{\scshape EPTA, InPTA:} collaboration, \emph{{The second data release from the European Pulsar Timing Array - III. Search for gravitational wave signals}}, \href{https://doi.org/10.1051/0004-6361/202346844}{\emph{Astron. Astrophys.} {\bfseries 678} (2023) A50} [\href{https://arxiv.org/abs/2306.16214}{{\ttfamily 2306.16214}}].

\bibitem{Reardon:2023gzh}
D.J.~Reardon et~al., \emph{{Search for an Isotropic Gravitational-wave Background with the Parkes Pulsar Timing Array}}, \href{https://doi.org/10.3847/2041-8213/acdd02}{\emph{Astrophys. J. Lett.} {\bfseries 951} (2023) L6} [\href{https://arxiv.org/abs/2306.16215}{{\ttfamily 2306.16215}}].

\bibitem{Miles:2024seg}
M.T.~Miles et~al., \emph{{The MeerKAT Pulsar Timing Array: The first search for gravitational waves with the MeerKAT radio telescope}}, \href{https://doi.org/10.1093/mnras/stae2571}{\emph{Mon. Not. Roy. Astron. Soc.} {\bfseries 536} (2025) 1489} [\href{https://arxiv.org/abs/2412.01153}{{\ttfamily 2412.01153}}].

\bibitem{LIGOScientific:2016jlg}
{\scshape LIGO Scientific, Virgo} collaboration, \emph{{Upper Limits on the Stochastic Gravitational-Wave Background from Advanced LIGO\textquoteright{}s First Observing Run}}, \href{https://doi.org/10.1103/PhysRevLett.118.121101}{\emph{Phys. Rev. Lett.} {\bfseries 118} (2017) 121101} [\href{https://arxiv.org/abs/1612.02029}{{\ttfamily 1612.02029}}].

\bibitem{TianQin:2015yph}
{\scshape TianQin} collaboration, \emph{{TianQin: a space-borne gravitational wave detector}}, \href{https://doi.org/10.1088/0264-9381/33/3/035010}{\emph{Class. Quant. Grav.} {\bfseries 33} (2016) 035010} [\href{https://arxiv.org/abs/1512.02076}{{\ttfamily 1512.02076}}].

\bibitem{Hu:2017mde}
W.-R.~Hu and Y.-L.~Wu, \emph{{The Taiji Program in Space for gravitational wave physics and the nature of gravity}}, \href{https://doi.org/10.1093/nsr/nwx116}{\emph{Natl. Sci. Rev.} {\bfseries 4} (2017) 685}.

\bibitem{Ruan:2018tsw}
W.-H.~Ruan, Z.-K.~Guo, R.-G.~Cai and Y.-Z.~Zhang, \emph{{Taiji program: Gravitational-wave sources}}, \href{https://doi.org/10.1142/S0217751X2050075X}{\emph{Int. J. Mod. Phys. A} {\bfseries 35} (2020) 2050075} [\href{https://arxiv.org/abs/1807.09495}{{\ttfamily 1807.09495}}].

\bibitem{Kawamura:2011zz}
S.~Kawamura et~al., \emph{{The Japanese space gravitational wave antenna: DECIGO}}, \href{https://doi.org/10.1088/0264-9381/28/9/094011}{\emph{Class. Quant. Grav.} {\bfseries 28} (2011) 094011}.

\bibitem{Maggiore:2019uih}
{\scshape ET} collaboration, \emph{{Science Case for the Einstein Telescope}}, \href{https://doi.org/10.1088/1475-7516/2020/03/050}{\emph{JCAP} {\bfseries 03} (2020) 050} [\href{https://arxiv.org/abs/1912.02622}{{\ttfamily 1912.02622}}].

\bibitem{Boyanovsky:2006bf}
D.~Boyanovsky, H.J.~de~Vega and D.J.~Schwarz, \emph{{Phase transitions in the early and the present universe}}, \href{https://doi.org/10.1146/annurev.nucl.56.080805.140539}{\emph{Ann. Rev. Nucl. Part. Sci.} {\bfseries 56} (2006) 441} [\href{https://arxiv.org/abs/hep-ph/0602002}{{\ttfamily hep-ph/0602002}}].

\bibitem{Mazumdar:2018dfl}
A.~Mazumdar and G.~White, \emph{{Review of cosmic phase transitions: their significance and experimental signatures}}, \href{https://doi.org/10.1088/1361-6633/ab1f55}{\emph{Rept. Prog. Phys.} {\bfseries 82} (2019) 076901} [\href{https://arxiv.org/abs/1811.01948}{{\ttfamily 1811.01948}}].

\bibitem{Athron:2023xlk}
P.~Athron, C.~Bal\'azs, A.~Fowlie, L.~Morris and L.~Wu, \emph{{Cosmological phase transitions: From perturbative particle physics to gravitational waves}}, \href{https://doi.org/10.1016/j.ppnp.2023.104094}{\emph{Prog. Part. Nucl. Phys.} {\bfseries 135} (2024) 104094} [\href{https://arxiv.org/abs/2305.02357}{{\ttfamily 2305.02357}}].

\bibitem{Caprini:2015zlo}
C.~Caprini et~al., \emph{{Science with the space-based interferometer eLISA. II: Gravitational waves from cosmological phase transitions}}, \href{https://doi.org/10.1088/1475-7516/2016/04/001}{\emph{JCAP} {\bfseries 04} (2016) 001} [\href{https://arxiv.org/abs/1512.06239}{{\ttfamily 1512.06239}}].

\bibitem{Caprini:2019egz}
C.~Caprini et~al., \emph{{Detecting gravitational waves from cosmological phase transitions with LISA: an update}}, \href{https://doi.org/10.1088/1475-7516/2020/03/024}{\emph{JCAP} {\bfseries 03} (2020) 024} [\href{https://arxiv.org/abs/1910.13125}{{\ttfamily 1910.13125}}].

\bibitem{Friedrich:2022cak}
L.S.~Friedrich, M.J.~Ramsey-Musolf, T.V.I.~Tenkanen and V.Q.~Tran, \emph{{Addressing the Gravitational Wave - Collider Inverse Problem}},  \href{https://arxiv.org/abs/2203.05889}{{\ttfamily 2203.05889}}.

\bibitem{Freitas:2021yng}
F.F.~Freitas, G.~Louren\c{c}o, A.P.~Morais, A.~Nunes, J.a.~Ol\'\i{}via, R.~Pasechnik et~al., \emph{{Impact of SM parameters and of the vacua of the Higgs potential in gravitational waves detection}}, \href{https://doi.org/10.1088/1475-7516/2022/03/046}{\emph{JCAP} {\bfseries 03} (2022) 046} [\href{https://arxiv.org/abs/2108.12810}{{\ttfamily 2108.12810}}].

\bibitem{Cline:2021iff}
J.M.~Cline, A.~Friedlander, D.-M.~He, K.~Kainulainen, B.~Laurent and D.~Tucker-Smith, \emph{{Baryogenesis and gravity waves from a UV-completed electroweak phase transition}}, \href{https://doi.org/10.1103/PhysRevD.103.123529}{\emph{Phys. Rev. D} {\bfseries 103} (2021) 123529} [\href{https://arxiv.org/abs/2102.12490}{{\ttfamily 2102.12490}}].

\bibitem{Paul:2020wbz}
A.~Paul, U.~Mukhopadhyay and D.~Majumdar, \emph{{Gravitational Wave Signatures from Domain Wall and Strong First-Order Phase Transitions in a Two Complex Scalar extension of the Standard Model}}, \href{https://doi.org/10.1007/JHEP05(2021)223}{\emph{JHEP} {\bfseries 05} (2021) 223} [\href{https://arxiv.org/abs/2010.03439}{{\ttfamily 2010.03439}}].

\bibitem{Niemi:2020hto}
L.~Niemi, M.J.~Ramsey-Musolf, T.V.I.~Tenkanen and D.J.~Weir, \emph{{Thermodynamics of a Two-Step Electroweak Phase Transition}}, \href{https://doi.org/10.1103/PhysRevLett.126.171802}{\emph{Phys. Rev. Lett.} {\bfseries 126} (2021) 171802} [\href{https://arxiv.org/abs/2005.11332}{{\ttfamily 2005.11332}}].

\bibitem{Wang:2019pet}
X.~Wang, F.P.~Huang and X.~Zhang, \emph{{Gravitational wave and collider signals in complex two-Higgs doublet model with dynamical CP-violation at finite temperature}}, \href{https://doi.org/10.1103/PhysRevD.101.015015}{\emph{Phys. Rev. D} {\bfseries 101} (2020) 015015} [\href{https://arxiv.org/abs/1909.02978}{{\ttfamily 1909.02978}}].

\bibitem{Hashino:2018wee}
K.~Hashino, R.~Jinno, M.~Kakizaki, S.~Kanemura, T.~Takahashi and M.~Takimoto, \emph{{Selecting models of first-order phase transitions using the synergy between collider and gravitational-wave experiments}}, \href{https://doi.org/10.1103/PhysRevD.99.075011}{\emph{Phys. Rev. D} {\bfseries 99} (2019) 075011} [\href{https://arxiv.org/abs/1809.04994}{{\ttfamily 1809.04994}}].

\bibitem{Tian:2024ysd}
C.~Tian, X.~Wang and C.~Bal\'azs, \emph{{Gravitational waves from cosmological first-order phase transitions with precise hydrodynamics}},  \href{https://arxiv.org/abs/2409.14505}{{\ttfamily 2409.14505}}.

\bibitem{Wang:2024slx}
X.~Wang, C.~Tian and C.~Bal\'azs, \emph{{Self-consistent prediction of gravitational waves from cosmological phase transitions}},  \href{https://arxiv.org/abs/2409.06599}{{\ttfamily 2409.06599}}.

\bibitem{Hindmarsh:2013xza}
M.~Hindmarsh, S.J.~Huber, K.~Rummukainen and D.J.~Weir, \emph{{Gravitational waves from the sound of a first order phase transition}}, \href{https://doi.org/10.1103/PhysRevLett.112.041301}{\emph{Phys. Rev. Lett.} {\bfseries 112} (2014) 041301} [\href{https://arxiv.org/abs/1304.2433}{{\ttfamily 1304.2433}}].

\bibitem{Hindmarsh:2015qta}
M.~Hindmarsh, S.J.~Huber, K.~Rummukainen and D.J.~Weir, \emph{{Numerical simulations of acoustically generated gravitational waves at a first order phase transition}}, \href{https://doi.org/10.1103/PhysRevD.92.123009}{\emph{Phys. Rev. D} {\bfseries 92} (2015) 123009} [\href{https://arxiv.org/abs/1504.03291}{{\ttfamily 1504.03291}}].

\bibitem{Hindmarsh:2017gnf}
M.~Hindmarsh, S.J.~Huber, K.~Rummukainen and D.J.~Weir, \emph{{Shape of the acoustic gravitational wave power spectrum from a first order phase transition}}, \href{https://doi.org/10.1103/PhysRevD.96.103520}{\emph{Phys. Rev. D} {\bfseries 96} (2017) 103520} [\href{https://arxiv.org/abs/1704.05871}{{\ttfamily 1704.05871}}].

\bibitem{Hindmarsh:2016lnk}
M.~Hindmarsh, \emph{{Sound shell model for acoustic gravitational wave production at a first-order phase transition in the early Universe}}, \href{https://doi.org/10.1103/PhysRevLett.120.071301}{\emph{Phys. Rev. Lett.} {\bfseries 120} (2018) 071301} [\href{https://arxiv.org/abs/1608.04735}{{\ttfamily 1608.04735}}].

\bibitem{Hindmarsh:2019phv}
M.~Hindmarsh and M.~Hijazi, \emph{{Gravitational waves from first order cosmological phase transitions in the Sound Shell Model}}, \href{https://doi.org/10.1088/1475-7516/2019/12/062}{\emph{JCAP} {\bfseries 12} (2019) 062} [\href{https://arxiv.org/abs/1909.10040}{{\ttfamily 1909.10040}}].

\bibitem{Guo:2020grp}
H.-K.~Guo, K.~Sinha, D.~Vagie and G.~White, \emph{{Phase Transitions in an Expanding Universe: Stochastic Gravitational Waves in Standard and Non-Standard Histories}}, \href{https://doi.org/10.1088/1475-7516/2021/01/001}{\emph{JCAP} {\bfseries 01} (2021) 001} [\href{https://arxiv.org/abs/2007.08537}{{\ttfamily 2007.08537}}].

\bibitem{Wang:2021dwl}
X.~Wang, F.P.~Huang and Y.~Li, \emph{{Sound velocity effects on the phase transition gravitational wave spectrum in the sound shell model}}, \href{https://doi.org/10.1103/PhysRevD.105.103513}{\emph{Phys. Rev. D} {\bfseries 105} (2022) 103513} [\href{https://arxiv.org/abs/2112.14650}{{\ttfamily 2112.14650}}].

\bibitem{Cai:2023guc}
R.-G.~Cai, S.-J.~Wang and Z.-Y.~Yuwen, \emph{{Hydrodynamic sound shell model}}, \href{https://doi.org/10.1103/PhysRevD.108.L021502}{\emph{Phys. Rev. D} {\bfseries 108} (2023) L021502} [\href{https://arxiv.org/abs/2305.00074}{{\ttfamily 2305.00074}}].

\bibitem{RoperPol:2023dzg}
A.~Roper~Pol, S.~Procacci and C.~Caprini, \emph{{Characterization of the gravitational wave spectrum from sound waves within the sound shell model}}, \href{https://doi.org/10.1103/PhysRevD.109.063531}{\emph{Phys. Rev. D} {\bfseries 109} (2024) 063531} [\href{https://arxiv.org/abs/2308.12943}{{\ttfamily 2308.12943}}].

\bibitem{Giombi:2024kju}
L.~Giombi, J.~Dahl and M.~Hindmarsh, \emph{{Signatures of the speed of sound on the gravitational wave power spectrum from sound waves}},  \href{https://arxiv.org/abs/2409.01426}{{\ttfamily 2409.01426}}.

\bibitem{Jinno:2022mie}
R.~Jinno, T.~Konstandin, H.~Rubira and I.~Stomberg, \emph{{Higgsless simulations of cosmological phase transitions and gravitational waves}}, \href{https://doi.org/10.1088/1475-7516/2023/02/011}{\emph{JCAP} {\bfseries 02} (2023) 011} [\href{https://arxiv.org/abs/2209.04369}{{\ttfamily 2209.04369}}].

\bibitem{Caprini:2019pxz}
C.~Caprini, D.G.~Figueroa, R.~Flauger, G.~Nardini, M.~Peloso, M.~Pieroni et~al., \emph{{Reconstructing the spectral shape of a stochastic gravitational wave background with LISA}}, \href{https://doi.org/10.1088/1475-7516/2019/11/017}{\emph{JCAP} {\bfseries 11} (2019) 017} [\href{https://arxiv.org/abs/1906.09244}{{\ttfamily 1906.09244}}].

\bibitem{Flauger:2020qyi}
R.~Flauger, N.~Karnesis, G.~Nardini, M.~Pieroni, A.~Ricciardone and J.~Torrado, \emph{{Improved reconstruction of a stochastic gravitational wave background with LISA}}, \href{https://doi.org/10.1088/1475-7516/2021/01/059}{\emph{JCAP} {\bfseries 01} (2021) 059} [\href{https://arxiv.org/abs/2009.11845}{{\ttfamily 2009.11845}}].

\bibitem{Giese:2021dnw}
F.~Giese, T.~Konstandin and J.~van~de Vis, \emph{{Finding sound shells in LISA mock data using likelihood sampling}}, \href{https://doi.org/10.1088/1475-7516/2021/11/002}{\emph{JCAP} {\bfseries 11} (2021) 002} [\href{https://arxiv.org/abs/2107.06275}{{\ttfamily 2107.06275}}].

\bibitem{Gowling:2022pzb}
C.~Gowling, M.~Hindmarsh, D.C.~Hooper and J.~Torrado, \emph{{Reconstructing physical parameters from template gravitational wave spectra at LISA: first order phase transitions}}, \href{https://doi.org/10.1088/1475-7516/2023/04/061}{\emph{JCAP} {\bfseries 04} (2023) 061} [\href{https://arxiv.org/abs/2209.13551}{{\ttfamily 2209.13551}}].

\bibitem{Guo:2024gmu}
H.-k.~Guo, F.~Hajkarim, K.~Sinha, G.~White and Y.~Xiao, \emph{{A Precise Fitting Formula for Gravitational Wave Spectra from Phase Transitions}},  \href{https://arxiv.org/abs/2407.02580}{{\ttfamily 2407.02580}}.

\bibitem{cybenko1989approximation}
G.~Cybenko, \emph{Approximation by superpositions of a sigmoidal function}, {\emph{Mathematics of control, signals and systems} {\bfseries 2} (1989) 303}.

\bibitem{HORNIK1989359}
K.~Hornik, M.~Stinchcombe and H.~White, \emph{Multilayer feedforward networks are universal approximators}, \href{https://doi.org/https://doi.org/10.1016/0893-6080(89)90020-8}{\emph{Neural Networks} {\bfseries 2} (1989) 359}.

\bibitem{HORNIK1991251}
K.~Hornik, \emph{Approximation capabilities of multilayer feedforward networks}, \href{https://doi.org/https://doi.org/10.1016/0893-6080(91)90009-T}{\emph{Neural Networks} {\bfseries 4} (1991) 251}.

\bibitem{LESHNO1993861}
M.~Leshno, V.Y.~Lin, A.~Pinkus and S.~Schocken, \emph{Multilayer feedforward networks with a nonpolynomial activation function can approximate any function}, \href{https://doi.org/https://doi.org/10.1016/S0893-6080(05)80131-5}{\emph{Neural Networks} {\bfseries 6} (1993) 861}.

\bibitem{Auld:2006pm}
T.~Auld, M.~Bridges, M.P.~Hobson and S.F.~Gull, \emph{{Fast cosmological parameter estimation using neural networks}}, \href{https://doi.org/10.1111/j.1745-3933.2006.00276.x}{\emph{Mon. Not. Roy. Astron. Soc.} {\bfseries 376} (2007) L11} [\href{https://arxiv.org/abs/astro-ph/0608174}{{\ttfamily astro-ph/0608174}}].

\bibitem{Auld:2007qz}
T.~Auld, M.~Bridges and M.P.~Hobson, \emph{{CosmoNet: Fast cosmological parameter estimation in non-flat models using neural networks}}, \href{https://doi.org/10.1111/j.1365-2966.2008.13279.x}{\emph{Mon. Not. Roy. Astron. Soc.} {\bfseries 387} (2008) 1575} [\href{https://arxiv.org/abs/astro-ph/0703445}{{\ttfamily astro-ph/0703445}}].

\bibitem{SpurioMancini:2021ppk}
A.~Spurio~Mancini, D.~Piras, J.~Alsing, B.~Joachimi and M.P.~Hobson, \emph{{CosmoPower: emulating cosmological power spectra for accelerated Bayesian inference from next-generation surveys}}, \href{https://doi.org/10.1093/mnras/stac064}{\emph{Mon. Not. Roy. Astron. Soc.} {\bfseries 511} (2022) 1771} [\href{https://arxiv.org/abs/2106.03846}{{\ttfamily 2106.03846}}].

\bibitem{Pal:2022hpi}
S.~Pal, P.~Chanda and R.~Saha, \emph{{Estimation of the Full-sky Power Spectrum between Intermediate and Large Angular Scales from Partial-sky CMB Anisotropies Using an Artificial Neural Network}}, \href{https://doi.org/10.3847/1538-4357/acb4ee}{\emph{Astrophys. J.} {\bfseries 945} (2023) 77} [\href{https://arxiv.org/abs/2203.14060}{{\ttfamily 2203.14060}}].

\bibitem{Nygaard:2022wri}
A.~Nygaard, E.B.~Holm, S.~Hannestad and T.~Tram, \emph{{CONNECT: a neural network based framework for emulating cosmological observables and cosmological parameter inference}}, \href{https://doi.org/10.1088/1475-7516/2023/05/025}{\emph{JCAP} {\bfseries 05} (2023) 025} [\href{https://arxiv.org/abs/2205.15726}{{\ttfamily 2205.15726}}].

\bibitem{Gunther:2022pto}
S.~G\"unther, J.~Lesgourgues, G.~Samaras, N.~Sch\"oneberg, F.~Stadtmann, C.~Fidler et~al., \emph{{CosmicNet II: emulating extended cosmologies with efficient and accurate neural networks}}, \href{https://doi.org/10.1088/1475-7516/2022/11/035}{\emph{JCAP} {\bfseries 11} (2022) 035} [\href{https://arxiv.org/abs/2207.05707}{{\ttfamily 2207.05707}}].

\bibitem{Bonici:2023xjk}
M.~Bonici, F.~Bianchini and J.~Ruiz-Zapatero, \emph{{Capse.jl: efficient and auto-differentiable CMB power spectra emulation}},  \href{https://arxiv.org/abs/2307.14339}{{\ttfamily 2307.14339}}.

\bibitem{Agarwal:2012ew}
S.~Agarwal, F.B.~Abdalla, H.A.~Feldman, O.~Lahav and S.A.~Thomas, \emph{{PkANN - I. Non-linear matter power spectrum interpolation through artificial neural networks}}, \href{https://doi.org/10.1111/j.1365-2966.2012.21326.x}{\emph{Mon. Not. Roy. Astron. Soc.} {\bfseries 424} (2012) 1409} [\href{https://arxiv.org/abs/1203.1695}{{\ttfamily 1203.1695}}].

\bibitem{Agarwal:2013aea}
S.~Agarwal, F.B.~Abdalla, H.A.~Feldman, O.~Lahav and S.A.~Thomas, \emph{{pkann \textendash{} II. A non-linear matter power spectrum interpolator developed using artificial neural networks}}, \href{https://doi.org/10.1093/mnras/stu090}{\emph{Mon. Not. Roy. Astron. Soc.} {\bfseries 439} (2014) 2102} [\href{https://arxiv.org/abs/1312.2101}{{\ttfamily 1312.2101}}].

\bibitem{Manrique-Yus:2019hqc}
A.~Manrique-Yus and E.~Sellentin, \emph{{Euclid-era cosmology for everyone: neural net assisted MCMC sampling for the joint 3 \texttimes{} 2 likelihood}}, \href{https://doi.org/10.1093/mnras/stz3059}{\emph{Mon. Not. Roy. Astron. Soc.} {\bfseries 491} (2020) 2655} [\href{https://arxiv.org/abs/1907.05881}{{\ttfamily 1907.05881}}].

\bibitem{Angulo:2020vky}
R.E.~Angulo, M.~Zennaro, S.~Contreras, G.~Aric\`o, M.~Pellejero-Iba\~nez and J.~St\"ucker, \emph{{The BACCO simulation project: exploiting the full power of large-scale structure for cosmology}}, \href{https://doi.org/10.1093/mnras/stab2018}{\emph{Mon. Not. Roy. Astron. Soc.} {\bfseries 507} (2021) 5869} [\href{https://arxiv.org/abs/2004.06245}{{\ttfamily 2004.06245}}].

\bibitem{Arico:2021izc}
G.~Aric\`o, R.E.~Angulo and M.~Zennaro, \emph{{Accelerating Large-Scale-Structure data analyses by emulating Boltzmann solvers and Lagrangian Perturbation Theory}},  \href{https://arxiv.org/abs/2104.14568}{{\ttfamily 2104.14568}}.

\bibitem{DeRose:2021pqx}
J.~DeRose, S.-F.~Chen, M.~White and N.~Kokron, \emph{{Neural network acceleration of large-scale structure theory calculations}}, \href{https://doi.org/10.1088/1475-7516/2022/04/056}{\emph{JCAP} {\bfseries 04} (2022) 056} [\href{https://arxiv.org/abs/2112.05889}{{\ttfamily 2112.05889}}].

\bibitem{Bose:2022vwi}
B.~Bose, M.~Tsedrik, J.~Kennedy, L.~Lombriser, A.~Pourtsidou and A.~Taylor, \emph{{Fast and accurate predictions of the non-linear matter power spectrum for general models of Dark Energy and Modified Gravity}}, \href{https://doi.org/10.1093/mnras/stac3783}{\emph{Mon. Not. Roy. Astron. Soc.} {\bfseries 519} (2023) 4780} [\href{https://arxiv.org/abs/2210.01094}{{\ttfamily 2210.01094}}].

\bibitem{Bonici:2025ltp}
M.~Bonici, G.~D'Amico, J.~Bel and C.~Carbone, \emph{{Effort: a fast and differentiable emulator for the Effective Field Theory of the Large Scale Structure of the Universe}},  \href{https://arxiv.org/abs/2501.04639}{{\ttfamily 2501.04639}}.

\bibitem{Kern:2017ccn}
N.S.~Kern, A.~Liu, A.R.~Parsons, A.~Mesinger and B.~Greig, \emph{{Emulating Simulations of Cosmic Dawn for 21 cm Power Spectrum Constraints on Cosmology, Reionization, and X-Ray Heating}}, \href{https://doi.org/10.3847/1538-4357/aa8bb4}{\emph{Astrophys. J.} {\bfseries 848} (2017) 23} [\href{https://arxiv.org/abs/1705.04688}{{\ttfamily 1705.04688}}].

\bibitem{Schmit:2017pho}
C.J.~Schmit and J.R.~Pritchard, \emph{{Emulation of reionization simulations for Bayesian inference of astrophysics parameters using neural networks}}, \href{https://doi.org/10.1093/mnras/stx3292}{\emph{Mon. Not. Roy. Astron. Soc.} {\bfseries 475} (2018) 1213} [\href{https://arxiv.org/abs/1708.00011}{{\ttfamily 1708.00011}}].

\bibitem{Bevins:2021eah}
H.T.J.~Bevins, W.J.~Handley, A.~Fialkov, E.d.L.~Acedo and K.~Javid, \emph{{globalemu: a novel and robust approach for emulating the sky-averaged 21-cm signal from the cosmic dawn and epoch of reionization}}, \href{https://doi.org/10.1093/mnras/stab2737}{\emph{Mon. Not. Roy. Astron. Soc.} {\bfseries 508} (2021) 2923} [\href{https://arxiv.org/abs/2104.04336}{{\ttfamily 2104.04336}}].

\bibitem{Jinno:2020eqg}
R.~Jinno, T.~Konstandin and H.~Rubira, \emph{{A hybrid simulation of gravitational wave production in first-order phase transitions}}, \href{https://doi.org/10.1088/1475-7516/2021/04/014}{\emph{JCAP} {\bfseries 04} (2021) 014} [\href{https://arxiv.org/abs/2010.00971}{{\ttfamily 2010.00971}}].

\bibitem{Jinno:2021ury}
R.~Jinno, T.~Konstandin, H.~Rubira and J.~van~de Vis, \emph{{Effect of density fluctuations on gravitational wave production in first-order phase transitions}}, \href{https://doi.org/10.1088/1475-7516/2021/12/019}{\emph{JCAP} {\bfseries 12} (2021) 019} [\href{https://arxiv.org/abs/2108.11947}{{\ttfamily 2108.11947}}].

\bibitem{Blasi:2023rqi}
S.~Blasi, R.~Jinno, T.~Konstandin, H.~Rubira and I.~Stomberg, \emph{{Gravitational waves from defect-driven phase transitions: domain walls}}, \href{https://doi.org/10.1088/1475-7516/2023/10/051}{\emph{JCAP} {\bfseries 10} (2023) 051} [\href{https://arxiv.org/abs/2302.06952}{{\ttfamily 2302.06952}}].

\bibitem{Espinosa:2010hh}
J.R.~Espinosa, T.~Konstandin, J.M.~No and G.~Servant, \emph{{Energy Budget of Cosmological First-order Phase Transitions}}, \href{https://doi.org/10.1088/1475-7516/2010/06/028}{\emph{JCAP} {\bfseries 06} (2010) 028} [\href{https://arxiv.org/abs/1004.4187}{{\ttfamily 1004.4187}}].

\bibitem{Wang:2020nzm}
X.~Wang, F.P.~Huang and X.~Zhang, \emph{{Energy budget and the gravitational wave spectra beyond the bag model}}, \href{https://doi.org/10.1103/PhysRevD.103.103520}{\emph{Phys. Rev. D} {\bfseries 103} (2021) 103520} [\href{https://arxiv.org/abs/2010.13770}{{\ttfamily 2010.13770}}].

\bibitem{Moore:1995si}
G.D.~Moore and T.~Prokopec, \emph{{How fast can the wall move? A Study of the electroweak phase transition dynamics}}, \href{https://doi.org/10.1103/PhysRevD.52.7182}{\emph{Phys. Rev. D} {\bfseries 52} (1995) 7182} [\href{https://arxiv.org/abs/hep-ph/9506475}{{\ttfamily hep-ph/9506475}}].

\bibitem{Wang:2020zlf}
X.~Wang, F.P.~Huang and X.~Zhang, \emph{{Bubble wall velocity beyond leading-log approximation in electroweak phase transition}},  \href{https://arxiv.org/abs/2011.12903}{{\ttfamily 2011.12903}}.

\bibitem{Jiang:2022btc}
S.~Jiang, F.P.~Huang and X.~Wang, \emph{{Bubble wall velocity during electroweak phase transition in the inert doublet model}}, \href{https://doi.org/10.1103/PhysRevD.107.095005}{\emph{Phys. Rev. D} {\bfseries 107} (2023) 095005} [\href{https://arxiv.org/abs/2211.13142}{{\ttfamily 2211.13142}}].

\bibitem{Kingma2014AdamAM}
D.P.~Kingma and J.~Ba, \emph{Adam: A method for stochastic optimization}, {\emph{CoRR} {\bfseries abs/1412.6980} (2014) }.

\bibitem{Cutting:2019zws}
D.~Cutting, M.~Hindmarsh and D.J.~Weir, \emph{{Vorticity, kinetic energy, and suppressed gravitational wave production in strong first order phase transitions}}, \href{https://doi.org/10.1103/PhysRevLett.125.021302}{\emph{Phys. Rev. Lett.} {\bfseries 125} (2020) 021302} [\href{https://arxiv.org/abs/1906.00480}{{\ttfamily 1906.00480}}].

\bibitem{Gowling:2021gcy}
C.~Gowling and M.~Hindmarsh, \emph{{Observational prospects for phase transitions at LISA: Fisher matrix analysis}}, \href{https://doi.org/10.1088/1475-7516/2021/10/039}{\emph{JCAP} {\bfseries 10} (2021) 039} [\href{https://arxiv.org/abs/2106.05984}{{\ttfamily 2106.05984}}].

\bibitem{Adams:2013qma}
M.R.~Adams and N.J.~Cornish, \emph{{Detecting a Stochastic Gravitational Wave Background in the presence of a Galactic Foreground and Instrument Noise}}, \href{https://doi.org/10.1103/PhysRevD.89.022001}{\emph{Phys. Rev. D} {\bfseries 89} (2014) 022001} [\href{https://arxiv.org/abs/1307.4116}{{\ttfamily 1307.4116}}].

\bibitem{Hindmarsh:2024tt}
M.~Hindmarsh, D.C.~Hooper, T.~Minkkinen and D.J.~Weir, \emph{{Recovering a phase transition signal in simulated LISA data with a modulated galactic foreground}}, \href{https://doi.org/10.1088/1475-7516/2025/04/052}{\emph{JCAP} {\bfseries 04} (2025) 052} [\href{https://arxiv.org/abs/2406.04894}{{\ttfamily 2406.04894}}].

\bibitem{WMAP:2003pyh}
{\scshape WMAP} collaboration, \emph{{First year Wilkinson Microwave Anisotropy Probe (WMAP) observations: Parameter estimation methodology}}, \href{https://doi.org/10.1086/377335}{\emph{Astrophys. J. Suppl.} {\bfseries 148} (2003) 195} [\href{https://arxiv.org/abs/astro-ph/0302218}{{\ttfamily astro-ph/0302218}}].

\bibitem{hoffman2014no}
M.D.~Hoffman, A.~Gelman et~al., \emph{The no-u-turn sampler: adaptively setting path lengths in hamiltonian monte carlo.}, {\emph{J. Mach. Learn. Res.} {\bfseries 15} (2014) 1593}.

\bibitem{1987PhLB..195..216D}
S.~Duane, A.D.~Kennedy, B.J.~Pendleton and D.~Roweth, \emph{{Hybrid Monte Carlo}}, \href{https://doi.org/10.1016/0370-2693(87)91197-X}{\emph{Phys. Lett. B} {\bfseries 195} (1987) 216}.

\bibitem{rumelhart1986learning}
D.E.~Rumelhart, G.E.~Hinton and R.J.~Williams, \emph{Learning representations by back-propagating errors}, {\emph{nature} {\bfseries 323} (1986) 533}.

\bibitem{Feroz:2007kg}
F.~Feroz and M.P.~Hobson, \emph{{Multimodal nested sampling: an efficient and robust alternative to MCMC methods for astronomical data analysis}}, \href{https://doi.org/10.1111/j.1365-2966.2007.12353.x}{\emph{Mon. Not. Roy. Astron. Soc.} {\bfseries 384} (2008) 449} [\href{https://arxiv.org/abs/0704.3704}{{\ttfamily 0704.3704}}].

\bibitem{Giese:2020rtr}
F.~Giese, T.~Konstandin and J.~van~de Vis, \emph{{Model-independent energy budget of cosmological first-order phase transitions\textemdash{}A sound argument to go beyond the bag model}}, \href{https://doi.org/10.1088/1475-7516/2020/07/057}{\emph{JCAP} {\bfseries 07} (2020) 057} [\href{https://arxiv.org/abs/2004.06995}{{\ttfamily 2004.06995}}].

\bibitem{Giese:2020znk}
F.~Giese, T.~Konstandin, K.~Schmitz and J.~van~de Vis, \emph{{Model-independent energy budget for LISA}}, \href{https://doi.org/10.1088/1475-7516/2021/01/072}{\emph{JCAP} {\bfseries 01} (2021) 072} [\href{https://arxiv.org/abs/2010.09744}{{\ttfamily 2010.09744}}].

\bibitem{Wang:2023jto}
X.~Wang, C.~Tian and F.P.~Huang, \emph{{Model-dependent analysis method for energy budget of the cosmological first-order phase transition}}, \href{https://doi.org/10.1088/1475-7516/2023/07/006}{\emph{JCAP} {\bfseries 07} (2023) 006} [\href{https://arxiv.org/abs/2301.12328}{{\ttfamily 2301.12328}}].

\bibitem{pan2009survey}
S.J.~Pan and Q.~Yang, \emph{A survey on transfer learning}, {\emph{IEEE Transactions on knowledge and data engineering} {\bfseries 22} (2009) 1345}.

\bibitem{flax2020github}
J.~Heek, A.~Levskaya, A.~Oliver, M.~Ritter, B.~Rondepierre, A.~Steiner et~al., \emph{{F}lax: A neural network library and ecosystem for {JAX}},  2024.

\bibitem{jax2018github}
J.~Bradbury, R.~Frostig, P.~Hawkins, M.J.~Johnson, C.~Leary, D.~Maclaurin et~al., \emph{{JAX}: composable transformations of {P}ython+{N}um{P}y programs},  2018.

\bibitem{phan2019composable}
D.~Phan, N.~Pradhan and M.~Jankowiak, \emph{Composable effects for flexible and accelerated probabilistic programming in numpyro},  \href{https://arxiv.org/abs/1912.11554}{{\ttfamily 1912.11554}}.

\bibitem{bingham2019pyro}
E.~Bingham, J.P.~Chen, M.~Jankowiak, F.~Obermeyer, N.~Pradhan, T.~Karaletsos et~al., \emph{Pyro: Deep universal probabilistic programming}, {\emph{J. Mach. Learn. Res.} {\bfseries 20} (2019) 28:1}.

\bibitem{Kumar2019}
R.~Kumar, C.~Carroll, A.~Hartikainen and O.~Martin, \emph{Arviz a unified library for exploratory analysis of bayesian models in python}, \href{https://doi.org/10.21105/joss.01143}{\emph{Journal of Open Source Software} {\bfseries 4} (2019) 1143}.

\end{thebibliography}\endgroup

\end{document}